\documentclass[aps,superscriptaddress,floatfix,twocolumn,footinbib]{revtex4-1}
\usepackage{amssymb}
\usepackage{amsmath}
\usepackage{amsfonts}
\usepackage{appendix}
\usepackage{bm}
\usepackage{graphicx}
\usepackage{epsfig}
\usepackage{epstopdf}
\usepackage{balance}
\usepackage[dvipsnames]{xcolor}
\usepackage{calc}
\usepackage{natbib}
\usepackage[colorlinks,
            linkcolor=blue,
            anchorcolor=blue,
            citecolor=blue,
            urlcolor=blue]{hyperref}
\usepackage{lipsum}

\usepackage{color}
\usepackage{soul}

\begin{document}
\title{Making artificial $\textit{p}_{x,y}$-orbital honeycomb electron lattice on metal surface}

\author{Wen-Xuan Qiu}
\affiliation{School of Physics and Wuhan National High Magnetic field center,
Huazhong University of Science and Technology, Wuhan 430074,  China}
\author{Liang Ma}
\affiliation{School of Physics and Wuhan National High Magnetic field center,
Huazhong University of Science and Technology, Wuhan 430074,  China}
\author{Jing-Tao L\"u}
\email{jtlu@hust.edu.cn}
\affiliation{School of Physics and Wuhan National High Magnetic field center,
Huazhong University of Science and Technology, Wuhan 430074,  China}
\author{Jin-Hua Gao}
\email{jinhua@hust.edu.cn}
\affiliation{School of Physics and Wuhan National High Magnetic field center,
Huazhong University of Science and Technology, Wuhan 430074,  China}

\begin{abstract}
We theoretically demonstrate that the desired $p_{x,y}$-orbital honeycomb electron lattice can be readily realized by arranging CO molecules into a hexagonal lattice on Cu(111) surface with scanning tunneling microscopy (STM). The electronic structure of the Cu surface states in the presence of CO molecules is calculated with various methods, \textit{i.e.}~DFT simulations, muffin-tin potential model and tight-binding model. Our calculations indicate that, by measuring the LDOS pattern using STM, the $p$-orbital surface bands can be immediately identified in experiment. We also give an analytic interpretation of the $p$-orbital LDOS pattern with $k \cdot p$ method. Meanwhile, different from the case of graphene, the $p$-orbital honeycomb lattice has two kinds of edge states, which can also be directly observed in STM experiment.
 Our work points out a feasible way to construct a  $p_{x,y}$-orbital honeycomb electron lattice in a real system, which may have exotic properties, such as Wigner crystal, ferromagnetism, $f$-wave superconductivity, quantum anomalous Hall (QAH) effect. Furthermore,   we also propose a simple way to calculate and identify the modified Cu surface bands in the Cu/CO systems with the DFT simulations.
Considering the recent works about $p$-orbital square lattice in similar systems [M. R. Slot, \textit{et al.} Nat. Phys. \textbf{13}, 672 (2017); Liang Ma, \textit{et al.} Phys.
Rev. B \textbf{99}, 205403 (2019)],  our work once again illustrates that the artificial electron lattice on metal surface is an ideal platform to study the orbital physics in a controllable way.
\end{abstract}

\maketitle
\section{introduction}
Orbital is an independent degree of freedom of electrons in solid in addition to their charge and spin. In many transition metal oxides, orbital physics plays an important role, and the interplay between the orbital, spin and charge degree of freedom can induce many exotic phenomena such as colossal magnetoresistance~\cite{PhysRevLett.77.175,ramirez1997colossal,PhysRevB.62.15096,PhysRevLett.93.227202}, superconductivity~\cite{Bednorz1986,luke1998time,ohtomo2004high,ja063355c}, and metal-insulator transition~\cite{PhysRevLett.90.066403,chen2007orbital,kong2008orbital}. Though the orbital physics is important, the study about the influence of orbital degree of freedom in real materials is still a big challenge, because orbital is always coupled with other degrees of freedom such as charge, spin, or crystal field.
Artificial lattice systems, $\textit{e.g.}$ cold atoms in optical lattice~\cite{PhysRevLett.99.200405}, photonic lattice~\cite{orbitaledgestate}, offer  ideal platforms to investigate the orbital  physics due to their unprecedented controllability. In last decade,  great efforts have been made to simulate the orbital physics in optical lattices~\cite{wu2009unconventional,li2016physics}. In experiment, bosons in  \textit{p}-band optical lattice have been realized and intensively studied~\cite{wirth2011evidence,olschlager2013interaction,PhysRevLett.114.115301}.

\begin{figure}[ht]
\centering
\includegraphics[width=0.5\textwidth,trim=0 0 0 0,clip]{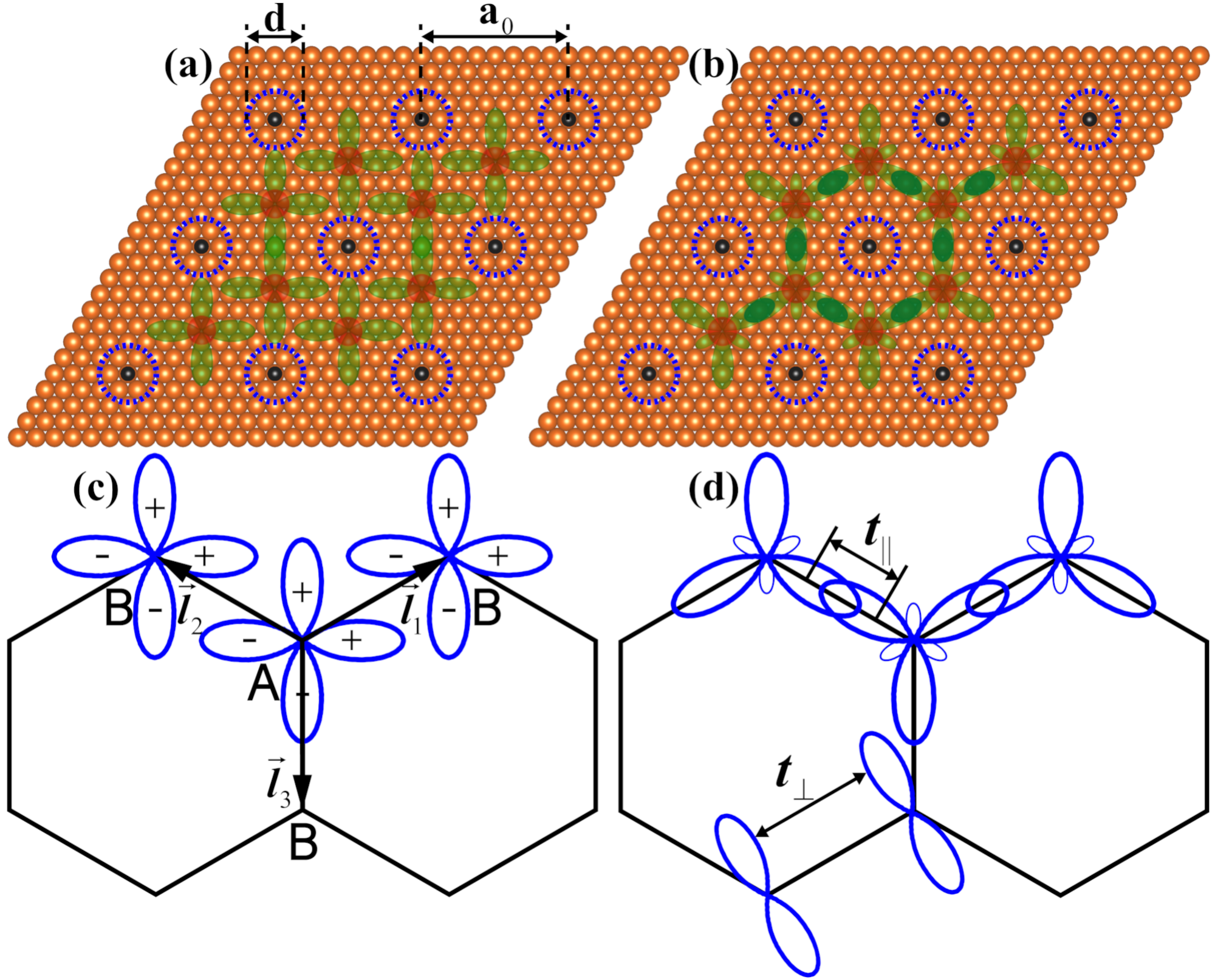}
\caption{(Color online). (a) and (b):  Schematic of the $p_{x,y}$-orbital honeycomb lattice on Cu(111) surface. CO molecules are denoted as black balls, blue circles represent the repulsive muffin-tin potential applied by CO molecules. $d$ is the diameter of the muffin-tin potential, $a_0$ is the lattice constant of this artificial honeycomb lattice.  Red discs denote the $s$ orbitals. In (a), green lobes represent the $p_{x,y}$ orbitals. In (b), we use green lobes to illustrate the $\sigma$ bonds formed by $p$ orbitals. (c) and (d): Schematic of the TB model of this $p_{x,y}$-orbital honeycomb lattice. In (d), $t_{\parallel}$ is the hopping between the projected $p$ orbitals on neighboring sites parallel to the bond direction; $t_{\perp}$ is the hopping between the projected $p$ orbitals perpendicular to the bond direction.
}
\label{fig1}
\end{figure}

Interestingly, some exotic orbital related quantum states, which do not exist in real materials, can be realized in artificial lattice systems. An intriguing example is the $\textit{p}_{x,y}$-orbital honeycomb lattice, which was proposed by Wu and coworkers in 2007~\cite{wu2007,wu2008}. Different from graphene, in optical lattice, the energy of  $s$ orbital is rather largely separated away from the $p$ orbital, so that a $\textit{p}$-orbital honeycomb lattice can be constructed without $sp$ hybridization.  The  $\textit{p}$-orbital honeycomb lattice has a unique band structure. It has four $p$ bands, where two of them are flat bands and the other two give rise to a Dirac cone at the $K$ points in Brillouin zone. Due to the quenched kinetic energy of flat bands,  strong correlation effects, such as Wigner crystal~\cite{wu2007}, ferromagnetism~\cite{PhysRevA.82.053618}, may appear. Meanwhile, the band structure is able to be tuned into topological nontrivial state~\cite{Zhang2019,ZhuGB2019}, \textit{e.g.}~the QAH effect~\cite{PhysRevLett.101.186807,PhysRevA.83.023615,PhysRevB.90.075114,Bloch2019,XieXC2019}and QSH effect~\cite{Ligang2018,Canonico2019}. It also enables unconventional $f$-wave Cooper pairing with a conventional attractive interaction~\cite{PhysRevA.82.053611}. Beyond the bulk properties, this $p_{x,y}$-orbital honeycomb lattice has two kinds of edge states, which are distinct from that of graphene. One is zero energy edge states, which have a similar origin to the conventional edge states in graphene. The other is novel dispersive edge states. Both of them have been observed in photonic lattice in a recent experiment~\cite{orbitaledgestate}.

However, fermions in $\textit{p}$-orbital honeycomb lattice, which have been intensively studied in theory, have not been reported in experiment. Very recently, it was also theoretically proposed that the $p_{x,y}$-orbital honeycomb lattice can be founded in some special materials~\cite{zhou2016,barreteau2017,song2018,jiang2019,zhou2019,HU2020}, but these proposals have not been confirmed in experiment so far.
To the best of our knowledge, experimental realization of this interesting $p$-orbital honeycomb lattice has only been reported in photonic lattice of coupled micropillars~\cite{orbitaledgestate, amo2014}. Recently, we propose that $p$-orbital bands can be realized in the artificial electron lattice on metal surface~\cite{ma2017orbital}, where the metal surface electrons are transformed into an electron lattice by periodically arranged adatoms. And the $p$-band features on square and Lieb electron lattice
have already been observed in a recent scanning tunneling microscopy (STM) experiment~\cite{slot2017experimental, ma2017orbital}.

In this work, we illustrate that the interesting $p_{x,y}$-orbital honeycomb electron lattice can be readily constructed on metal surface with the same technique by designing a proper adatom pattern. And it actually has been realized in  experiment~\cite{gomes2012designer,wang2014manipulation}, only further measurement is needed to give a confirmation.  The scheme is illustrated in Fig.~\ref{fig1}. Here,  the CO molecules are periodically deposited on Cu(111) surface with the help of the STM tip. This Cu/CO system has been successfully used to construct the artificial honeycomb~\cite{park2009making,gomes2012designer,wang2014manipulation}, square~\cite{slot2017experimental}, Lieb electron lattices~\cite{slot2017experimental} on metal surface. In Fig.~\ref{fig1} (a),  CO molecules are represented by black balls, where each applies a repulsive potential on metal surface. The repulsive potential can be approximately described as a muffin-tin potential, which is denoted as the blue circles in Fig.~\ref{fig1} (a). As pointed out in Ref.~\cite{park2009making,gomes2012designer,wang2014manipulation}, if the CO molecules are arranged into a hexagonal lattice (Fig.~\ref{fig1}), the Cu surface states can be forced into a honeycomb electron lattice because of the repulsive potential applied by the CO molecules. Former works~\cite{park2009making,gomes2012designer,wang2014manipulation} only consider the lowest two energy bands, which correspond to the $s$ orbital and has a graphene like band structure. We would like to point out that, in the same system, the higher bands are from the artificial $p$ orbitals, which is just the desired $p_{x,y}$-orbital honeycomb electron lattice [Fig.~\ref{fig1} (a),(c)].
As shown in Fig.~\ref{fig1} (b) and (d),  the $p_{x,y}$ orbitals will form $\sigma$ bonds parallel to the bond direction, and $\pi$ bonds perpendicular to the bond direction as well~\cite{wu2007,wu2008}.
The LDOS of the $p$ orbitals can be directly observed by STM.
We use density functional theory (DFT) simulations, muffin-tin potential model, tight-binding (TB) model and $k \cdot p$ method to interpret this $p$-orbital picture. We also illustrate that the two kinds of edge states of this $p$-orbital honeycomb lattice could be directly observed in this artificial electron lattice system.

To calculate the modified Cu(111) surface bands in the presence of CO adatoms, the muffin-tin model is the most common method, while the DFT simulations do not work well so far~\cite{ropo2014density,Coexist2016}. The reason is that, in DFT simulations, we have to consider a large supercell, including a thick Cu slab and CO adatoms,  in which the surface bands and the Cu bulk bands are heavily folded and thoroughly mixed together. Thus, the major obstacle for the DFT simulations is to distinguish the modified surface bands from massive bulk bands. Here, we illustrate that the modified Cu surface bands can be identified by proper layer resolved band structure projections, so that now the desired Cu surface bands in the Cu/CO systems can be directly calculated using the DFT simulations. It should be a general method to calculate the adatom modified metal surface states via DFT simulations.

The outline of this paper is as follows: in Sec.~\ref{sec1}, we give the models and methods used in the calculations; in Sec.~\ref{sec2}, we show the numerical results and the corresponding discussions; finally, a short summary is given in Sec~\ref{sec3}.

\section{model and methods}\label{sec1}
Here, we introduce the theoretical methods, \textit{i.e.}~muffin-tin potential model, tight-binding model and DFT simulations, which we used here to calculate the modified Cu surface bands in the Cu/CO systems.
\begin{figure}[ht]
\centering
\includegraphics[width=0.50\textwidth,trim=0 0 0 0,clip]{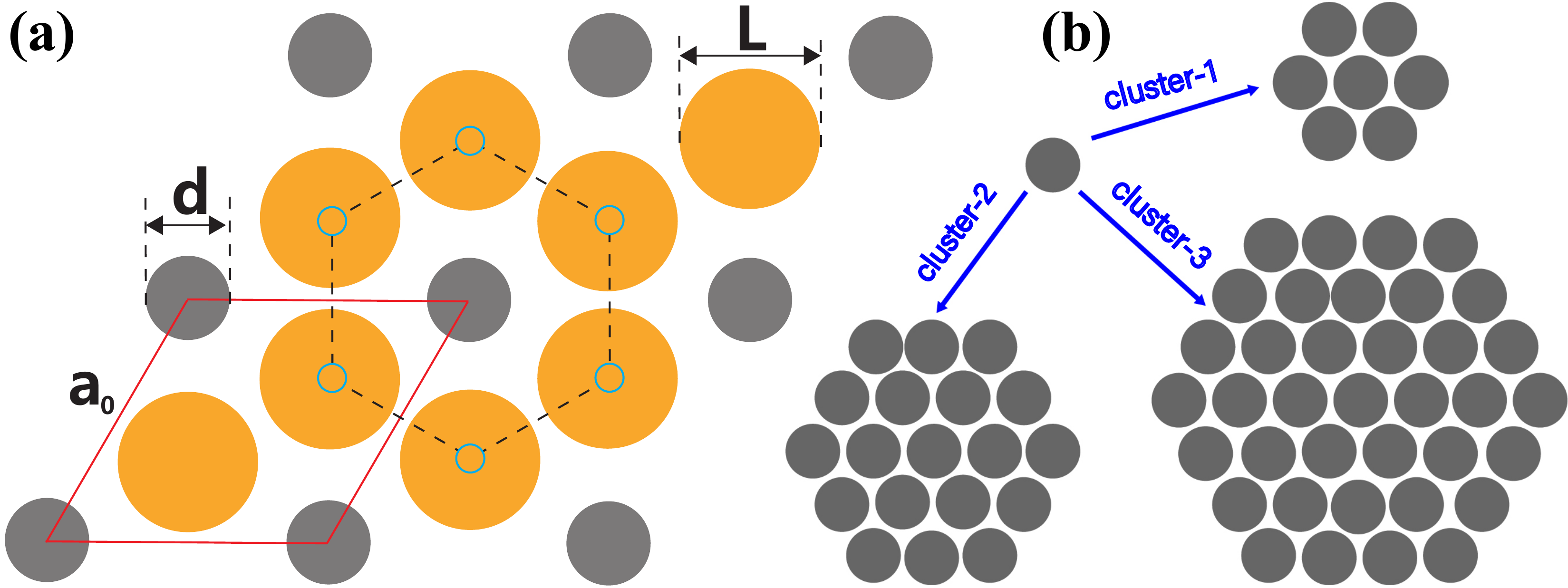}
\caption{(Color online). (a) Schematic of the muffin-tin potential model for this artificial honeycomb electron lattice. The repulsive potential of each CO molecule is represented as a black disc, which is $U_\textrm{CO}>0$ inside the disc and zero elsewhere, and  $d$ is its diameter. The lattice sites of honeycomb lattice are denoted as the orange discs, $a_0$ is the lattice constant. Instead of the muffin-tin potential model, we can equivalently use a cylindrical potential well (orange disc, $L$ is its diameter) to describe the ``artificial atom", by which the surface electrons are confined around the lattice sites. (b) Each CO adatom can be replaced by a CO cluster with different size in order to apply a  stronger potential on the Cu surface states.}
\label{fig2}
\end{figure}

\subsection{Muffin-tin potential model}
As mentioned above, the Cu surface states in  Cu/CO system can be intuitively described by a muffin-tin potential model. We illustrate this muffin-tin potential model in Fig.~\ref{fig2} (a).
 Here, CO molecules can be approximately viewed as a repulsive muffin-tin potential $U(r)$ (black discs), which is $U_{\textrm{CO}}>0$ inside the black discs and zero elsewhere. Thus, the surface electrons are confined into the regions in between CO molecules, which actually gives rise to a honeycomb lattice. The sites of the honeycomb lattice are represented by the orange discs in Fig.~\ref{fig2} (a). The Hamiltonian of the muffin-tin model is
\begin{equation}\label{Hmuffin}
    H_{Cu} = -\frac{\hbar^2}{2m^{*}}\nabla^2 + U(r).
\end{equation}
This Hamiltonian can be solved with plane wave method to get the energy bands of Cu surface states in the presence of CO molecules~\cite{park2009making,li2016designing,qiu2016designing}. Here, $m^*=0.38m_0$ is the effective mass of Cu surface states,  where $m_0$ is the electron mass~\cite{gomes2012designer}. The parameters of the muffin-tin potential $U(r)$ for the Cu/CO system are $U_{\textrm{CO}}=$ 0.8 eV and the diameter $d=$ 0.5 nm,
which are got by fitting the DFT results, as will be shown later.

The muffin-tin potential of a single CO adatom has a fixed value. However, we can get a stronger effective potential applied on Cu surface via replacing the single CO adatom with a CO cluster~\cite{slot2018experimental}. For example, several kinds of CO clusters with different sizes are plotted in Fig.~\ref{fig2} (b). We use cluster-$l$ to label the CO cluster where the side length of hexagon is $l$ times the length of the Cu-Cu distance. Note that the larger the cluster size is, the stronger influence of the effective potential is.

The LDOS can be got by
\begin{equation}\label{mtldos}
\begin{split}
\textrm{LDOS}(r,\varepsilon) = \sum_{mk\sigma}{|\phi_{mk\sigma}(r)|}^2\delta(\varepsilon-\varepsilon_{mk\sigma}).
\end{split}
\end{equation}
We will show that the low energy bands got by the muffin-tin potential model can be well interpreted as the $s$ and $p$ bands of a  honeycomb lattice.

\subsection{Orbitals in artificial atom}
We then discuss the concept of orbitals in the artificial electron lattice. It should be noted that each site (orange disc) of this honeycomb lattice in Fig.~\ref{fig2} (a) actually can be viewed as an artificial atom. In each artificial atom, the electrons are confined in the region enclosed by the adjacent CO molecules. Considering this potential confinement,  the eigenstates of an isolated artificial atom are discrete, which are very similar to the orbitals of a real atom. Then, through hopping between these artificial orbitals, we get the energy bands of this artificial electron lattice.

However, this orbital picture (and TB model) does not always work. This is because that the confinement potential is finite. If the electron kinetic energy is large enough, surface electrons can not be confined. Thus, the surface states should be described by the nearly free electron model instead of the TB model. In this situation, the concept of artificial orbitals is invalid.

We can roughly use a  cylindrical potential well $U_{atom}$ to describe this two dimensional artificial atom.  $H_{atom}=-\frac{\hbar^2}{2m^{*}}{\nabla}^{2} + U_{atom}(\textbf{r})$,
where
\begin{equation}
U_{atom}(\textbf{r})=\left\{
\begin{split}
-U_0, \qquad |\textbf{r}| \leq \frac{L}{2}\\
0, \qquad |\textbf{r}| > \frac{L}{2}
\end{split}
\right.
\end{equation}
$U_0$ and L are the value and diameter of the potential well, respectively.  One artificial atom corresponds to one cylindrical potential well, which is represented by one orange disc in Fig.~\ref{fig2} (a). The electrons trapped in $U_{atom}$ form the orbitals, and the hopping between orbitals gives rise to TB bands.

The orbitals can be got by calculating the bound states in the potential well.  The eigenfunction of $H_{atom}$ is
\begin{equation}
\phi(r,\theta) = \textrm{R}(r) \textrm{Y}(\theta),
\end{equation}
where $r$ is the radial coordinate and $\theta$ is the polar angle.
We have
\begin{equation}\label{angle}
\begin{split}
\frac{d^2\textrm{Y}(\theta)}{d\theta^2} = -n^2 \textrm{Y}(\theta) \qquad n=0, 1, 2, \cdots
\end{split}
\end{equation}
where, $\textrm{Y}_n(\theta)= a_n \cos(n\theta) + b_n \sin(n\theta) $ for $n>0$, and  $Y_0(\theta)=1/\sqrt{2\pi}$. Here, $a_n$ and $b_n$ are the coefficients, which can take on any value (but the wavefunction should be normalized).
The equations of radial part are
\begin{equation}\label{radial}
\left\{
\begin{split}
r^2\frac{d^2\textrm{R}(r)}{dr^2} + r\frac{d\textrm{R}(r)}{dr} + (\lambda_1 r^2 - n^2)\textrm{R}(r)=0, \qquad r\leq\frac{L}{2}\\
r^2\frac{d^2\textrm{R}(r)}{dr^2} + r\frac{d\textrm{R}(r)}{dr} - (\lambda_2 r^2 + n^2)\textrm{R}(r)=0, \qquad r>\frac{L}{2}
\end{split}
\right.
\end{equation}
where $\lambda_1=\frac{2m(\varepsilon+U_0)}{\hbar^2}$, $\lambda_2=-\frac{2m\varepsilon}{\hbar^2}$.
At the boundary $r_0=\frac{L}{2}$, the continuity of the wavefunction and its derivative should be satisfied,
\begin{equation}
\left\{
\begin{split}
&c_1 J_n(\sqrt{\lambda_1}r_0) = c_2 K_n(\sqrt{\lambda_2}r_0)\\
&c_1 J^{'}_{n}(\sqrt{\lambda_1}r_0)\sqrt{\lambda_1} = c_2 K^{'}_{n}(\sqrt{\lambda_2}r_0)\sqrt{\lambda_2}\\
&|c_1|^2\int_{0}^{r_0}r{J_n}^2(\sqrt{\lambda_1}r)dr + |c_2|^2\int_{r_0}^{\infty}r{K_n}^2(\sqrt{\lambda_2}r)dr = 1
\end{split}
\right.
\end{equation}
Here, the last equation is the requirement of wave function normalization, $c_1$ and $c_2$ are two coefficients to be determined. $J_n$ ($K_n$) is the Bessel (Hankel) function. Together with this boundary condition, we can get the eigenvalues and eigenfunctions of the bound states in this potential well.

For example,  in the case of $U_0=0.9$ eV and $L=1.8$ nm, there is one bound state with $n=0$, \textit{i.e.} the $s$ orbital. With $n=1$, the allowed $p$ orbitals are
\begin{equation}\label{porbital}
\begin{split}
\phi_{p_x}(r,\theta)=\frac{1}{\sqrt{\pi}}\textrm{R}(r)\cos{\theta}\\
\phi_{p_y}(r,\theta)=\frac{1}{\sqrt{\pi}}\textrm{R}(r)\sin{\theta}
\end{split}
\end{equation}
where
\begin{equation}
\textrm{R}(r)=\left\{
\begin{split}
c_1 \textbf{J}_1(\sqrt{\lambda_1}r), \qquad r\leq\frac{L}{2}\\
c_2 \textbf{K}_1(\sqrt{\lambda_2}r), \qquad r>\frac{L}{2}
\end{split}
\right.
\end{equation}
The coefficients are $c_1 \approx 2.20$, $c_2 \approx 1.16$, and the energy of the $p_{x,y}$ orbitals is $\varepsilon_{p_x}=\varepsilon_{p_y}=\varepsilon \approx -0.074$ eV. However, no solution exists for $n>1$, which implies that only $s$ and $p$ orbitals are valid in this situation. Note that, the larger $U_0$ is, the more  artificial orbitals are allowed.

The value of $U_{atom}$ for the Cu/CO system can be estimated by fitting the bands calculated from the muffin-tin model. For the case  in the experiment~\cite{gomes2012designer}, a reasonable $U_{atom}$ is: $U_0=0.9$ eV, $L=1.8$ nm, which is just the above example. So, we argue that only $s$ and $p$ orbitals are valid in such Cu/CO systems.

We illustrate the artificial orbitals in the honeycomb lattice in Fig.~\ref{fig1} (a). Red discs represent the $s$ orbitals, and green lobes denote the $p_{x,y}$ orbitals. Different from graphene, the $s$ orbitals are far away from the $p$ orbitals in energy, so that there is no $sp$ hybridization here. The $s$  orbitals give rise to an artificial graphene, which has already been observed in experiment~\cite{gomes2012designer,wang2014manipulation}. In the following, we focus on  the $p$ orbitals. Note that in two dimensional electron system, there are only $p_{x,y}$ orbitals, and no $p_z$ orbital. Thus, we actually get a $p_{x,y}$-orbital honeycomb lattice,  which is illustrated in Fig.~\ref{fig1} (a) and (c). The honeycomb lattice has two sublattices, A and B. So, we have four $p$ bands above the two $s$ bands in this artificial honeycomb lattice.

Note that the cylindrical potential well is a rough approximation of the confinement potential applied by CO molecules, which is just used to illustrate the concept of artificial orbitals. As shown in Fig.~\ref{fig2} (a), the regions where the surface electrons are confined are actually anisotropic and have three-fold rotational symmetry. Such anisotropy of the confinement potential is not included in the cylindrical (TB) model. This does not obviously influence the $s$ bands, but will give rise to a difference between the cylindrical (TB) band structure of $p$ orbitals and that got from muffin-tin model.

\subsection{Tight-binding model}\label{sectb}
The tight-binding  model of  $p_{x,y}$-orbital honeycomb lattice is given in Ref.~\cite{wu2008}. Here, we give a short introduction to this TB model.
First, we define three vectors $\Vec{l}_{1,2,3}$ [see in Fig.~\ref{fig1} (c)], which point from A site to three nearest neighbor B sites, i.e., the directions of bonds.  $a_0$ is the lattice constant [see in Fig.~\ref{fig1} (a)].  Considering the lattice geometry, $p_{x,y}$ orbitals are projected to directions parallel to $\Vec{l}_{i=1,2,3}$ forming the projected orbital $p_i=(p_x \Vec{e}_x + p_y \Vec{e}_y)\cdot \frac{\sqrt{3} \Vec{l}_i}{a_0}$, where $\Vec{e}_x$ ($\Vec{e}_y$) is the unit vector along x (y) direction~\cite{wu2008}. We also define $p'_{i=1,2,3}$ as the projected $p$ orbitals perpendicular to $\Vec{l}_{i=1,2,3}$. Note that here only two of $p_i$ or $p'_i$ orbitals are linearly independent. Bonds exist between these projected orbitals. There are two kinds of bonds between the $p$ orbitals on neighboring sites, \textit{i.e.}, $\sigma$ and $\pi$ bond. As shown in Fig.~\ref{fig1} (d), $\sigma$ bond corresponds to the hopping between $p_i$ orbitals (``head to tail") , and $\pi$ bond corresponds to the hopping between $p'_i$ orbitals (``shoulder by shoulder"). The Hamiltonian is $H=H_\sigma + H_\pi$, where
\begin{equation}
    H_\sigma = t_{\parallel} \sum_{\Vec{r}\in A,i} \{p^{\dagger}_{\Vec{r},i}p_{\Vec{r}+\hat{l}_i,i} + h.c.\},
\end{equation}
\begin{equation}
      H_\pi = -t_{\perp} \sum_{\Vec{r}\in A,i} \{p'^{\dagger}_{\Vec{r},i}p'_{\Vec{r}+\hat{l}_i,i} + h.c.\}.
\end{equation}
Here, $t_{\parallel}$ ($t_{\perp}$) is the hopping of $\sigma$ ($\pi$) bond. The Hamiltonian can be viewed as the summation of all $\sigma$ and $\pi$ bonds.

It is convenient to use the basis $[p^A_{x},p^A_{y},p^B_{x},p^B_{y}]$, the corresponding Hamiltonian matrix is,
\begin{equation}\label{hamiltonian}
\begin{aligned}
H(k)=t_{\parallel}\left(
\begin{array}{cc}
0&H_{\parallel}\\
H^{+}_{\parallel}&0\\
\end{array}
\right)-t_{\perp}\left(
\begin{array}{cc}
0&H_{\perp}\\
H^{+}_{\perp}&0\\
\end{array}
\right),
\end{aligned}
\end{equation}
where
\begin{equation}\label{hsgm}
\begin{aligned}
H_{\parallel}=\left(
\begin{array}{cc}
\frac{3}{4}(e^{i \textbf{k} \textbf{l}_1}+e^{i \textbf{k} \textbf{l}_2})&\frac{\sqrt{3}}{4}(e^{i \textbf{k} \textbf{l}_1}-e^{i \textbf{k} \textbf{l}_2})\\
\frac{\sqrt{3}}{4}(e^{i \textbf{k} \textbf{l}_1}-e^{i \textbf{k} \textbf{l}_2})&\frac{1}{4}(e^{i \textbf{k} \textbf{l}_1}+e^{i \textbf{k} \textbf{l}_2})+e^{i \textbf{k} \textbf{l}_3}\\
\end{array}
\right),
\end{aligned}
\end{equation}

\begin{equation}
\begin{aligned}
H_{\perp}=\left(
\begin{array}{cc}
\frac{1}{4}(e^{i \textbf{k} \textbf{l}_1}+e^{i \textbf{k} \textbf{l}_2})+e^{i \textbf{k} \textbf{l}_3}&\frac{\sqrt{3}}{4}(e^{i \textbf{k} \textbf{l}_2}-e^{i \textbf{k} \textbf{l}_1})\\
\frac{\sqrt{3}}{4}(e^{i \textbf{k} \textbf{l}_2}-e^{i \textbf{k} \textbf{l}_1})&\frac{3}{4}(e^{i \textbf{k} \textbf{l}_1}+e^{i \textbf{k} \textbf{l}_2})\\
\end{array}
\right).
\end{aligned}
\end{equation}
Same expression can be obtained if we apply the Slater-Koster formalism~\cite{Slater-Koster} to $p_{x,y}$ orbitals. Diagonalizing $H(k)$, we can get the TB energy bands and wavefunctions.
Using the $p$-orbital basis in Eq.~\eqref{porbital}, we can further calculate the LDOS based on the TB model. The details to calculate the LDOS are given in Sec.~\ref{sec23}. Meanwhile, we can also get the edge states by calculating the bands of the ribbon structure based on this TB model~\cite{orbitaledgestate}.

\subsection{DFT simulations}\label{secdft}
In order to give more information about the experiment, DFT simulations can be done where more details of the Cu/CO system are included~\cite{dft2014,Coexist2016,ma2017orbital}.
As will be shown later, our DFT simulations are also in good agreement with the muffin-tin potential model, which means that the muffin-tin potential model captures the essential features of the experimental observations.

The DFT simulations use the Vienna Ab-initio Simulation Package (VASP), where projector-augmented wave method and a plane wave basis set are used~\cite{vasp}. We select the Perdew-Burke-Ernzerhof (PBE) version of the generalized gradient approximation~\cite{PBE}. The plane wave cutoff energy is $400$ eV and the $k$-point mesh is $9\times9\times1$. We use a Cu slab to mimic the Cu(111) surface, see in Fig.~\ref{fig1}. The lattice constant $a_0$ is the distance between two adjacent CO molecules. Here, $a_0$ is eight times the length of the Cu-Cu distance (about 2.04 nm), namely an $8\times8$ supercell is used. To avoid the interaction between nearest cells we set the inter-cell vacuum space to be $22{\textrm{\AA}}$. CO molecules are adsorbed on the top of Cu atoms, and  the Cu-C distance and C-O distance are set to be $1.85{\textrm{\AA}}$ and $1.155{\textrm{\AA}}$, respectively. We only optimize the positions of C, O atoms and the Cu atoms closest to C atoms. The convergence criteria for force acting on each atom is $0.02\textrm{eV}/{\textrm{\AA}}$. This has been shown to be accurate enough to capture the physics discussed here~\cite{ma2017orbital}.

We first calculate the energy bands of the Cu/CO systems, and then identify the modified surface bands via proper layer-resolved band projections. We also simulate the LDOS of the surface states and the corresponding STM image with the Tersoff-Hamann scheme~\cite{tersoff1983}. Finally, we calculate the charge difference caused by CO adsorption using the VASPKIT code~\cite{vei2019}, in order to study the effect of charge transfer between CO molecules and Cu slab.

\begin{figure*}[t]
\centering
\includegraphics[width=1.0\textwidth,trim=0 0 0 0,clip]{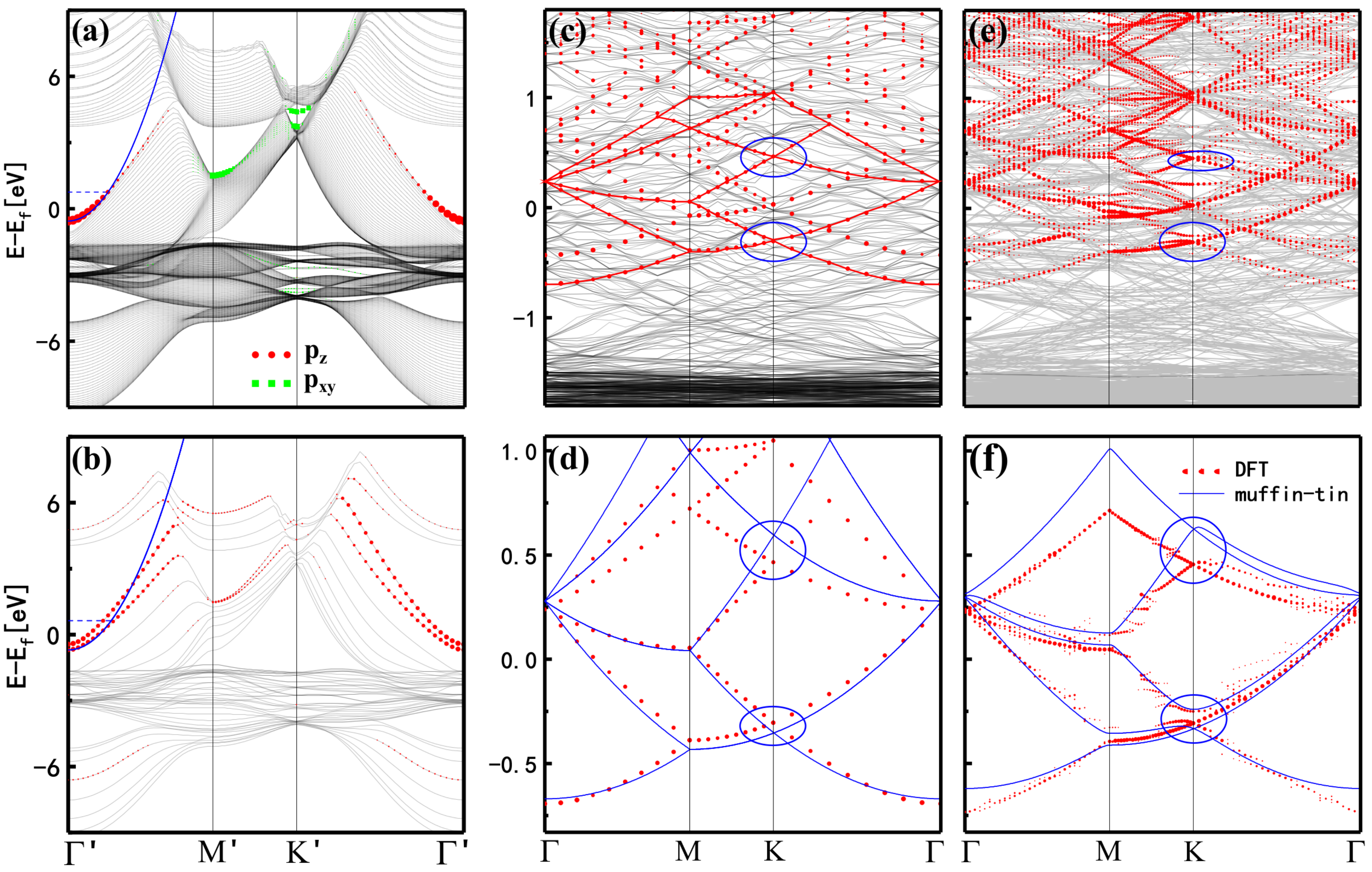}
\caption{(Color online). Energy bands got by DFT simulations and the corresponding orbital projections of Cu atoms in the top two layers. (a) and (b) are the energy bands of a pure Cu slab in an unfolded Brillouin zone. (a) 51-layer Cu slab. (b) 8-layer Cu slab. (c) and (e) are the energy bands of a 8-layer slab in a folded Brillouin zone when a $8\times8$ supercell is used. (c) Without CO adsorption. (e) With CO adsorption. (d) and (f) Fitting the lower energy surface band selected from (c) and (e) with the muffin-tin band. The fitting parameters are: $m^*=0.50m_0$, $U_{\textrm{CO}}=$ 0.8 eV and $d=$ 0.5 nm. Note that in order to save calculation resource, we only calculate 10 points in each $k$-path in (c) and (d). Other results are calculated with 30 points in each $k$-path.}
\label{fig3}
\end{figure*}

\section{Results and Discussions}\label{sec2}

\subsection{Energy bands}\label{sec21}
Various theoretical methods, \textit{i.e.}~DFT simulations, muffin-tin model and TB model, are used to calculate the modified Cu surface bands in the Cu/CO systems. The muffin-tin model and TB model can give an intuitive picture about the $p$ bands, which clearly interprets the  results of DFT simulations. Meanwhile, the parameters of the muffin-tin model and TB model can be got by fitting the results of DFT simulations.

\subsubsection{DFT simulations}
We first discuss how to calculate the Cu surface bands via DFT simulations. Here, we use a Cu slab to mimic the Cu surface. As mentioned above, the challenge here is to distinguish the surface bands from the bulk bands of the Cu slab. In the following, we use three steps to illustrate the Cu surface bands in the Cu/CO system.

At the first step,  we illustrate how to identify the Shockley surface bands of a Cu slab in the DFT simulations.
In Fig.~\ref{fig3} (a), we  plot the energy bands of a 51-layer-thick pure Cu slab, in which the surface bands are well known in former literature~\cite{butti2005image,berland2012response,courths2001from,kevan1983evidence,jeandupeux1999thermal,tamai2013spin}. As we see here, in such Cu slab, the surface bands are mixed up with the bulk bands. In order to identify the surface bands, we project all the bands of the Cu slab into the surface layers by layer resolved band structure projections.
The most important finding is that the Shockley surface states are mainly composed of the $p_z$ orbitals of surface layers, \textit{i.e.}, Cu $p_z$ orbitals of the top two layers, see the red dotted lines in Fig.~\ref{fig3} (a). Note that the red dotted lines here are just the surface bands reported in former literatures~\cite{butti2005image,berland2012response,courths2001from,kevan1983evidence,jeandupeux1999thermal,tamai2013spin}, and these states are strongly localized at the surface layers (See Fig.~\ref{figs1} (h) and (i) in Appendix~\ref{appb}).
The effective mass of the surface bands is about $m^*=0.45m_0$ (the blue solid line is the fitted $k^2$ dispersion).
Thus, we can use this property  to identify the surface bands in the DFT simulations.

A key fact here is that the Cu slab has two surfaces, which gives rise to two degenerate surface bands in the DFT simulations.   If the thickness of Cu slab is largely reduced, the two degenerate surface bands will be split due to the finite size effect.  In Fig.~\ref{fig3} (b), we plot the bands of an 8-layer-thick pure Cu slab. With the same method,  we get two split surface bands instead (red dotted lines), due to the coupling between the top and bottom surface states. The band split is about 0.2 eV. Such surface band split is unreal and does not exist in reality, since the sample of Cu(111) surface is thick enough. In principle, it can be avoided by using a thick Cu slab in the calculations. However, when we consider the CO adatoms, a very large supercell is needed for the Cu slab, so that the thickness of the Cu slab in the DFT simulations should be limited. For example, if we use an $8 \times 8$ supercell ($a_0=2.04$ nm) and an 8-layer-thick Cu(111) slab, we have 512 Cu atoms in one supercell which has already reached the limit of our calculation resource. Fortunately,  except for the band split,  such coupling only slightly changes the effective mass of the surface bands in this situation ($m^*=0.5 m_0$ and blue solid line is the fitted $k^2$ dispersion). It can still be viewed as the free electron band with $k^2$ dispersion. Therefore, in order to compare with experiment,  a reasonable and feasible way is to choose an 8-layer-thick Cu slab (maximumly allowable in our DFT simulations), and artificially select one of the two calculated surface bands.

At the second step, we show the folded surface bands where a supercell of the Cu slab is considered. The surface band of Cu(111) surface has a $k^2$ dispersion near the $\Gamma$ point, which can be approximately viewed as a free electron gas. But if the CO molecules are periodically adsorbed,  a supercell of the Cu slab has to be used as illustrated in Fig.~\ref{fig1} (a). In this situation, the surface bands are folded into a small BZ defined by the supercell. For example, Fig.~\ref{fig3} (c) gives the folded surface bands (red dots) of the 8-layer-thick Cu slab with an $8 \times 8$ supercell. Note that here we only consider a pure Cu(111) surface and CO molecules have not been included yet.  Since the 8-layer-thick Cu slab has two surface bands, we have two sets of folded surface bands here, which have an energy split about $0.2$ eV.  We select one set of the folded surface bands (low energy one), see the red solid lines in Fig.~\ref{fig3} (c). Such folded surface band calculated by DFT simulations can be well understood by the muffin-tin model in Eq.~\eqref{Hmuffin}. Here,  we set $U_{\textrm{CO}}=0$ in the muffin-tin model, since CO molecules have not been adsorbed yet. Thus, the muffin-tin model gives the folded free electron bands here.  In Fig.~\ref{fig3} (d), the red dots are the replotted folded surface bands got by DFT simulations, \textit{i.e.}, the red solid lines in Fig.~\ref{fig3} (c), and the blue solid lines are the folded surface bands got by the muffin-tin model ($m^*=0.50m_0$ and $U_{\textrm{CO}}=0$). We see that they coincide well with each other, especially in the low energy region.  We argue that the small discrepancy is because that the assumption of $k^2$ dispersion is invalid for the DFT surface bands at high energy, as shown in Fig.~\ref{fig3} (a) and  (b).

At the last step, we calculate the modified Cu surface bands in the presence of CO adatoms via DFT simulations.  In Fig.~\ref{fig3} (e), we show the bands of Cu/CO system, where an $8 \times 8$ supercell is used and CO adatoms are included as shown in Fig.~\ref{fig1} (a). Here, the red dots are the calculated surface bands via the layer projection method. Similarly, we select one of the two sets of surface bands, which are replotted in Fig.~\ref{fig3} (f). The blue lines are the surface bands calculated by muffin-tin model with $U_{\textrm{CO}}=0.8$ eV. The consistency of the DFT bands and that got from the muffin-tin model clearly indicates that the muffin-tin model gives a rather well description about the surface bands of the Cu/CO system. And here we also get the value of muffin-tin potential parameters, \textit{i.e.} $U_{\textrm{CO}}=0.8$ eV and $d=0.5$ nm,  by fitting the DFT bands.  It is clear to see the influence of the CO adatoms by comparing Fig.~\ref{fig3} (f) to (d). In Fig.~\ref{fig3} (f), the CO potential lifts the band degeneracy along some high symmetry lines, and a Dirac point at the lowest two bands emerges at $K$ point~\cite{park2009making,dvorn2012}, denoted by the blue circle. Actually, the lowest two bands here are just the $s$ bands of the honeycomb lattice, which give rise to an artificial graphene.  Above the lowest two bands, there is another Dirac point at $K$ point, also denoted by a blue circle.  As will be shown later, these high energy bands  above the $s$ bands, as well as the Dirac point,   just result from the $p_{x,y}$ orbitals of the honeycomb lattice.

\begin{figure}[t]
\centering
\includegraphics[width=0.5\textwidth,trim=0 0 0 0,clip]{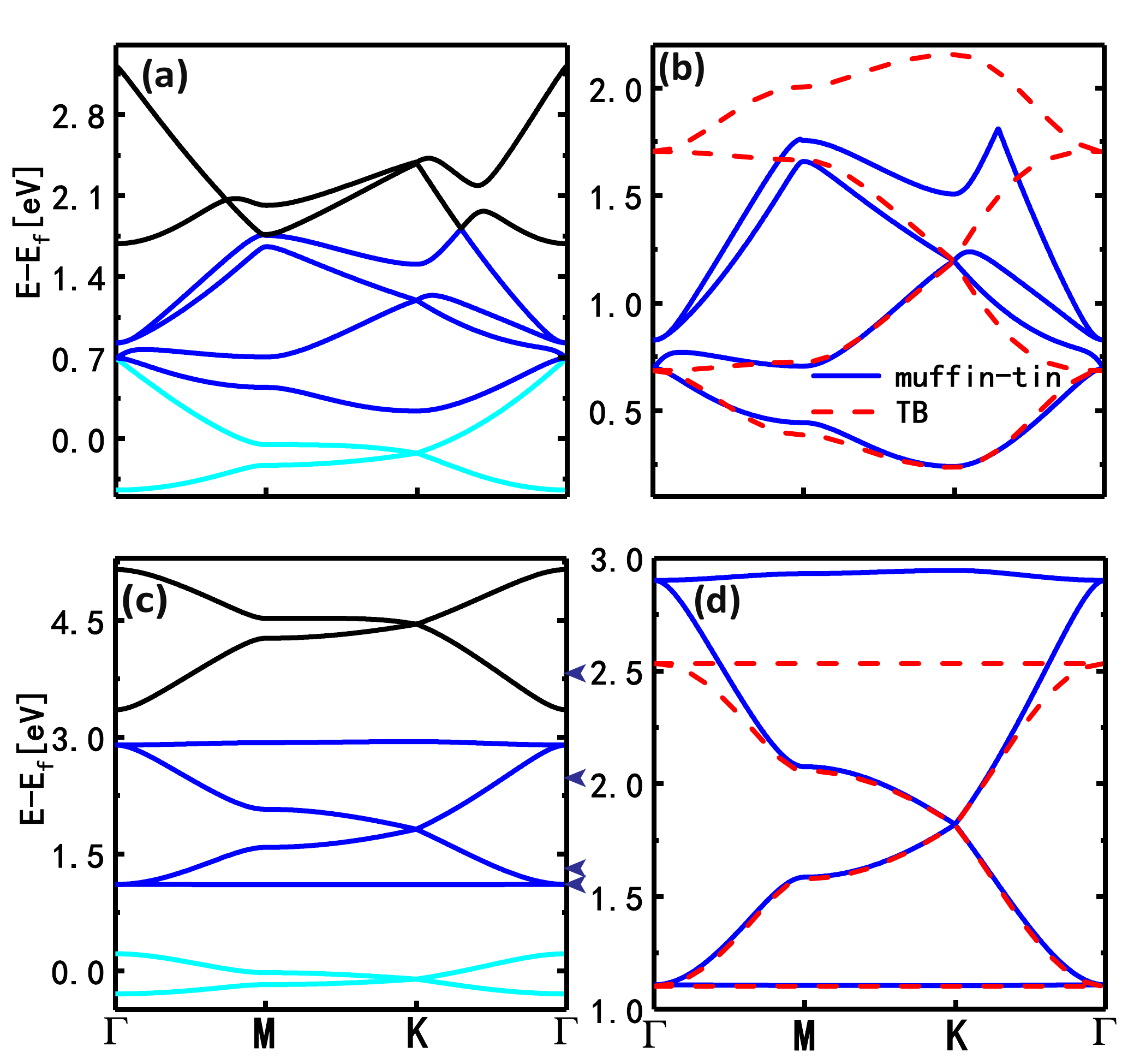}
\caption{(Color online). (a) Energy bands calculated by the muffin-tin potential model of single-CO case in Fig.~\ref{fig1} (a). $U_{\textrm{CO}}=9$ eV, $d=0.5$ nm, $a_0=2.04$ nm. Cyan lines are $s$ bands, blue lines are $p$ bands. (b) Fitting of the $p$ bands in (a) with $p$ orbital TB model. Red dashed lines are the TB bands. The fitting parameters are: $t_{\parallel}=0.49$ eV, $t_{\perp}=0.15$ eV, $\varepsilon_{p}=1.2$ eV. (c) Energy bands calculated by the muffin-tin potential model in the extreme case of Fig.~\ref{fig1} (a). $U_{\textrm{CO}}=9$ eV, $d=1.6$ nm, $a_0=2.04$ nm. Black lines here correspond to new $s$ orbitals. (d) Fitting of the $p$ bands in (c) with  TB model. The fitting parameters are: $t_{\parallel}=0.477$ eV, $t_{\perp}=0$ eV, $\varepsilon_{p}=1.82$ eV.}
\label{fig4}
\end{figure}
\subsubsection{Muffin-tin and TB models}\label{sec212}
Through DFT simulations, we have shown that the modified Cu surface bands in Cu/CO systems can be well described by the muffin-tin model. Here, based on the muffin-tin and TB models, we then illustrate that such Cu/CO systems do host a $p_{x,y}$ honeycomb electron lattice.

The surface bands calculated by muffin-tin model are plotted in Fig.~\ref{fig4} (a).  Here, to demonstrate,  we  use an ideal set of parameters for the muffin-tin model, $U_{\textrm{CO}}=9$ eV and $d=0.5$ nm.  As discussed in former literature~\cite{ma2017orbital,slot2018experimental},
 the lowest two bands (cyan lines) are from the $s$ orbitals, which are similar to the bands of graphene. This is just the reason why this artificial structure is named as ``artificial graphene"~\cite{park2009making}. We are interested in the upper four bands (blue lines). Actually, the four bands here are from the $p_{x,y}$ orbitals on the honeycomb lattice. To prove this point,   we use the TB model in Eq.~\eqref{hamiltonian} to fit the $p$-orbital energy bands in Fig.~\ref{fig4} (b). The red dashed lines are the results of TB model, and we also replot the $p$ bands from muffin-tin potential model (blue lines) as a comparison. We see that the $p$ bands got from the TB model coincide well with that from the muffin-tin potential model in the low energy region. The fitted parameters of the TB model are : $t_{\parallel}=0.49$ eV, $t_{\perp}=0.15$ eV and $\varepsilon_p=1.2$ eV.  At high energy, there is an obvious discrepancy. There are two reasons.
One is due to the anisotropy of the confinement potential as mentioned in the last section, which has been ignored in the TB model. It does not obviously influence the $s$ bands, but can not be completely ignored for the $p$ bands.  The other is due to hybridization between the $p$ orbitals and the states with higher energy.
Since the confine potential is finite, the surface states at higher energy here are more like that of nearly free electrons, for which the orbital picture and TB model do not work.

\begin{figure}[t]
\centering
\includegraphics[width=0.45\textwidth,trim=0 0 0 0,clip]{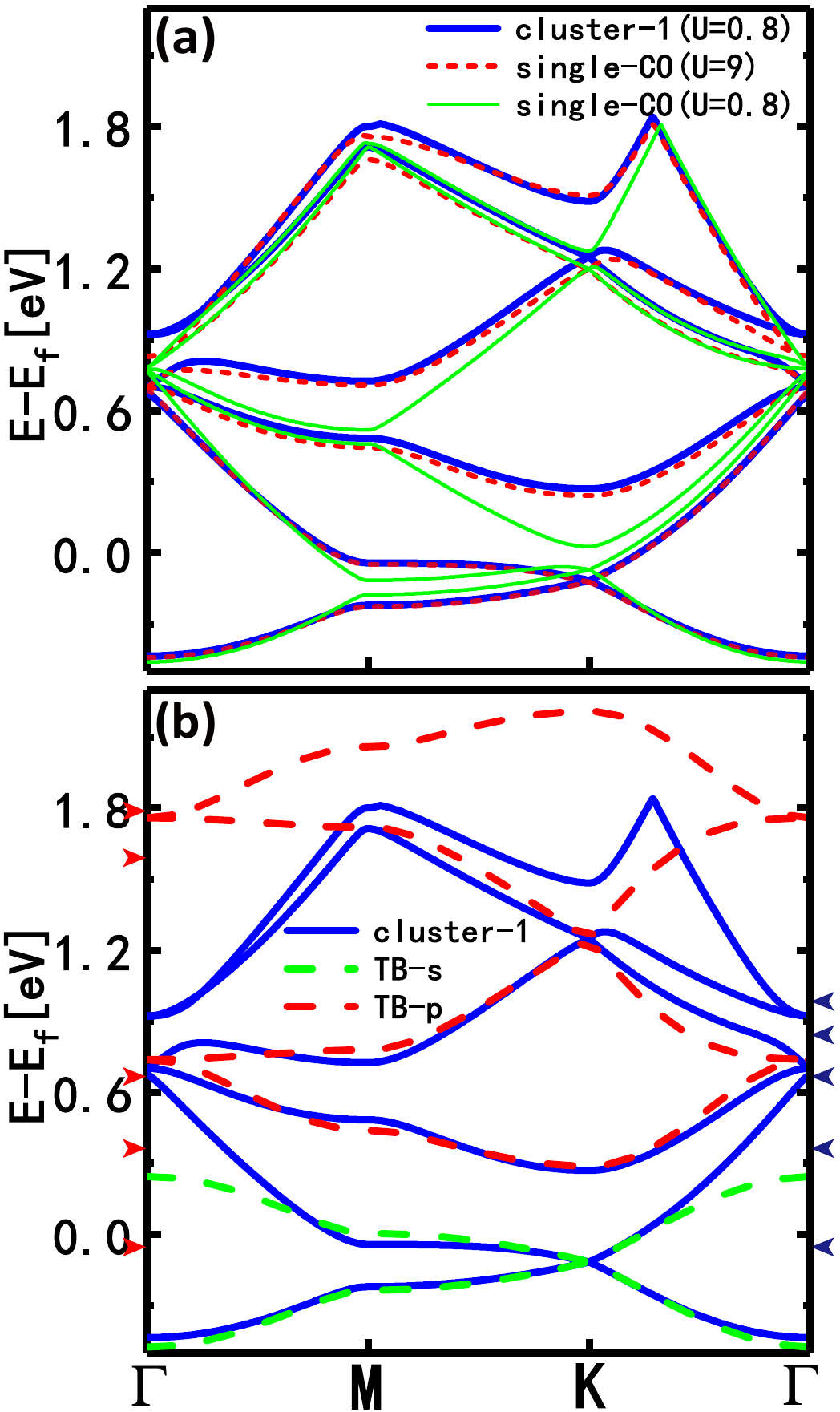}
\caption{(Color online). (a) Energy bands calculated by the muffin-tin potential model. Blue lines are the muffin-tin bands of cluster-1 with $U_{\textrm{CO}}=0.8$ eV, $d=0.5$ nm. Red dotted lines are the muffin-tin bands in Fig.~\ref{fig4} (a). Green lines are the muffin-tin bands of single-CO case in Fig.~\ref{fig1} (a) with $U_{\textrm{CO}}=0.8$ eV, $d=0.5$ nm. (b) Fitting of the bands (cluster-1) in (a) with TB model. Green dashed lines are the $s$ orbital TB bands with fitting parameters: $t_{s}=-0.12$ eV, $\varepsilon_{s}=-0.124$ eV; Red dashed lines are the $p$ orbital TB bands with fitting parameters: $t_{\parallel}=0.49$ eV, $t_{\perp}=0.15$ eV, $\varepsilon_{p}=1.248$ eV.}
\label{fig5}
\end{figure}

To make the $p$-orbital picture clearer, we consider an extreme case with $U_{\textrm{CO}}=9$ eV, $d=1.6$ nm. In this case, confinement potential is extremely large. Thus, more orbitals appear and these orbitals are well separated in energy. We plot the bands in this extreme case in Fig.~\ref{fig4} (c), (d). The lowest two bands are from $s$ orbitals. The upper four bands are just from the $p_{x,y}$ orbitals. Now, we see that there are two nearly flat bands and a Dirac point at the $\textrm{K}$ point, which are the characteristics of the bands of $p_{x,y}$-orbital honeycomb lattice~\cite{wu2007}. We also give a fitting with TB model, as shown in Fig.~\ref{fig4} (d).
Now, there is no hybridization with higher bands, and the discrepancy of the band structure at high energy should only result from the influence of the anisotropy of the confinement potential.

Interestingly, in Fig.~\ref{fig4} (c), the two bands above the $p$ bands are similar to the lowest two $s$ bands (black lines). Actually, they correspond to a new $s$ orbital of the artificial atom, which appears due to the very  large repulsive (confinement) potential in this extreme case.
So, we get another two graphene-like $s$ bands at high energy. However, they have distinct LDOS patterns, which will be shown later.

For the Cu/CO system,  we get $U_{\textrm{CO}}=0.8$ eV by fitting the DFT bands. However, the muffin-tin model tells that such repulsive potential is not strong enough to completely separate the $p$ bands from the $s$ bands in energy, and may not give a clear signal of $p$ band in experiment. In Fig.~\ref{fig5} (a), we plot the muffin-tin bands calculated with  $U_{\textrm{CO}}=0.8$ eV (green solid lines) and $U_{\textrm{CO}}=9$ eV (red dashed lines), respectively. We see that, with a weak repulsive potential, \textit{e.g.} $U_{\textrm{CO}}=0.8$ eV, the lowest $p$ band is badly bent, and obviously deviates from the typical shape of $p$ band. It implies that, with a single CO adatom,  the $p$ band features in experiment are not quite evident. A better way is to use a CO cluster instead of a single CO adatom, as shown in Fig.~\ref{fig2} (b), which has  been successfully utilized in experiment~\cite{slot2018experimental}.
The reason is that the CO cluster is essentially equivalent to a stronger effective potential in muffin-tin model. In Fig.~\ref{fig5} (a), we also plot the muffin-tin bands calculated with the cluster-1 of CO, see the blue solid lines. It obviously demonstrates that, with a CO cluster, we can get the same surface band structures as that from a strong repulsive potential (\textit{i.e.}, single adatom with $U_{\textrm{CO}}=9$ eV).
So, in order to  give a more clear  $p$ band characteristic, unless specified otherwise, we use the cluster-1 of CO in the muffin-tin model instead of a single CO adatom. This should be a more appropriate experimental design to observe the $p$ orbital bands in Cu/CO system.
We emphasize that  the physical picture of $p$ orbital remains the same no matter whether  a single CO adatom or a CO cluster is used.

In Fig.~\ref{fig5} (b), we use TB model to fit the muffin-tin  bands of the cluster-1 case.  The red (green) dashed lines are the TB bands of $p$ ($s$) orbitals.

\begin{figure*}[ht]
\centering
\includegraphics[width=1\textwidth,trim=0 0 0 0,clip]{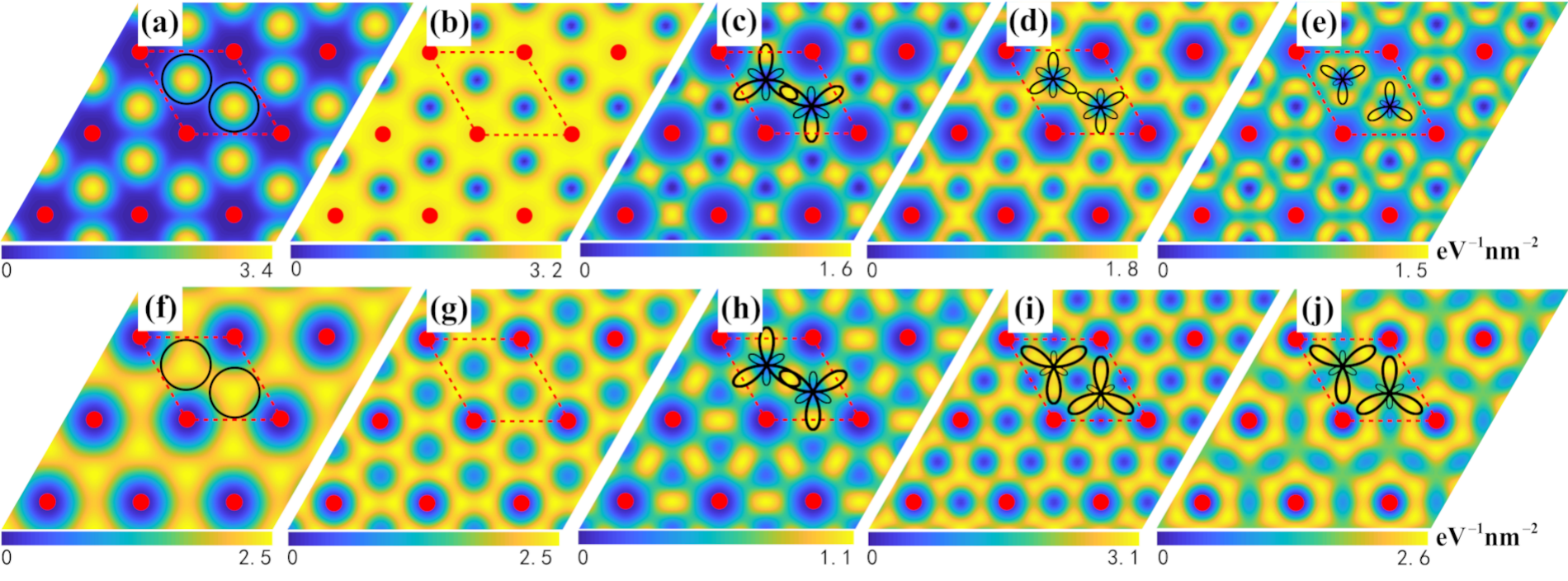}
\caption{(Color online). (a-e) LDOS patterns calculated with TB model at $E-E_f=$ -0.05, 0.36, 0.66, 1.59, 1.79 eV, respectively. Other parameters of this model are the same as that of TB bands in Fig.~\ref{fig5} (b). (f-j) LDOS patterns calculated with muffin-tin model at $E-E_f=$ -0.05, 0.36, 0.66, 0.84, 0.98 eV, respectively. Other parameters of this model are the same as that of muffin-tin bands in Fig.~\ref{fig5} (b).}
\label{fig6}
\end{figure*}
\subsection{LDOS pattern: numerical results}\label{sec22}
Now, we discuss the LDOS pattern in this artificial honeycomb lattice, which can be directly measured using STM. The observed LDOS pattern can be used to identify the $p$ bands in experiment.

Fig.~\ref{fig6} (a-e) are the LDOS calculated with the TB model at $E-E_f=$ -0.05, 0.36, 0.66, 1.59, 1.79 eV, respectively. The corresponding TB bands have been given in Fig.~\ref{fig5} (b),  and the energy of the LDOS are denoted by red arrows. In Fig.~\ref{fig6} (a), the LDOS corresponds to an energy at $s$ band ($E-E_f$= -0.05 eV). It clearly shows a honeycomb lattice pattern~\cite{park2009making,gomes2012designer}, in which the disc at each site represents an onsite $s$ orbital. Fig.~\ref{fig6} (b-c) are the LDOS of the bottom $p$-orbital flat band. Note that the flat band here is not exactly flat, so that the LDOS at different energy of this band will show distinct features. For example, Fig.~\ref{fig6} (b) gives the LDOS at energy around the K point, while Fig.~\ref{fig6} (c) is around the $\Gamma$ point. Near the $\Gamma$ point, the pattern shows a bonding state feature, in which the electron density mainly distributes in between the two lattice sites and is tiny near the center of the artificial atom. In contrast, near the K point the distribution becomes more complex and nonlocal, as shown in Fig.~\ref{fig6} (b). As for the middle two $p$ bands, we find two different kinds of LDOS patterns. At lower energy, the LDOS shows a bonding state feature quite like that in Fig.~\ref{fig6} (c). At higher energy, the LDOS has an anti-bonding state feature, which has a node in the middle of the bond, as shown in Fig.~\ref{fig6} (d). The LDOS of the top flat band has a stronger anti-bonding state feature, as shown in Fig.~\ref{fig6} (e), in which the electron density inside of the bond is pushed to the opposite directions outside of the bond. We will give a further analysis of the above LDOS patterns with $k \cdot p$ method later in Sec.~\ref{sec23}.

Fig.~\ref{fig6} (f-j) are the LDOS calculated with the muffin-tin potential model at $E-E_f=$ -0.05, 0.36, 0.66, 0.84, 0.98 eV, respectively. The corresponding band structures are also given in Fig.~\ref{fig5} (b), and the energy of the LDOS are denoted by blue arrows. Note that the $s$ bands and the bottom $p$-orbital flat band coincide well with that from the TB model. At higher energy, the $p$ bands have a discrepancy with that from the TB model, as shown in Fig.~\ref{fig5} (b). So it is reasonable to expect that the muffin-tin LDOS patterns have similar shapes as the TB LDOS patterns at low energy, while they are different at high energy. Fig.~\ref{fig6} (f) shows the LDOS of the $s$ bands and Fig.~\ref{fig6} (g-h) show the LDOS of the bottom $p$-orbital flat band near the K point and $\Gamma$ point. They are qualitatively in agreement with the TB results. When we move to the middle $p$ bands, only the anti-bonding features can be observed, as shown in Fig.~\ref{fig6} (i-j). We think that the muffin-tin potential model is more realistic. We expect that the LDOS patterns observed in experiment should be more like that from the muffin-tin potential model.

\begin{figure}[ht]
\centering
\includegraphics[width=0.5\textwidth,trim=0 0 0 0,clip]{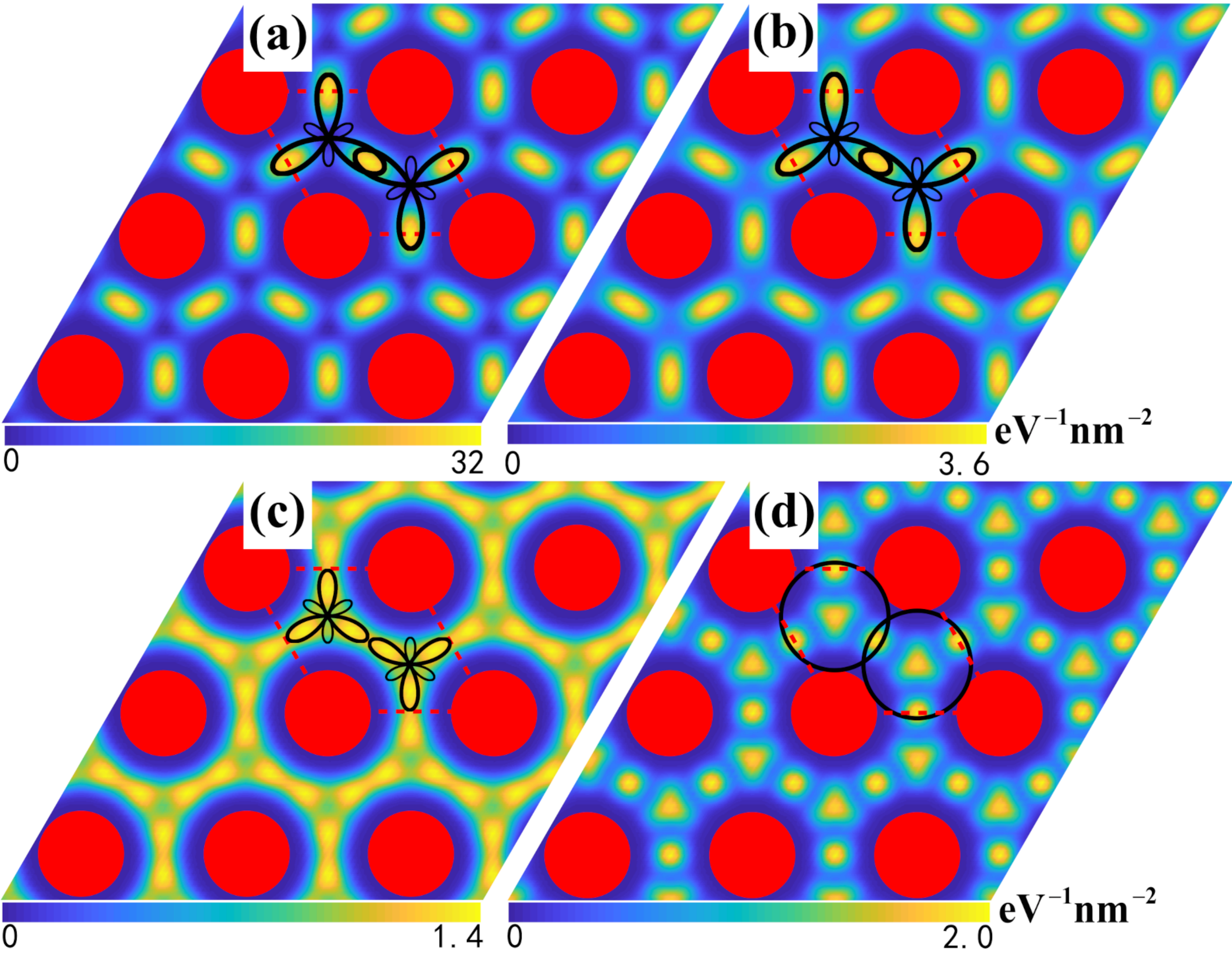}
\caption{(Color online). LDOS patterns in the artificial honeycomb lattice calculated with muffin-tin potential model for the extreme case in Fig.~\ref{fig4} (c). (a) $E-E_f=$ 1.11 eV, (b) $E-E_f=$ 1.32 eV, (c) $E-E_f=$ 2.47 eV, (d) $E-E_f=$3.82 eV. Other parameters are the same as Fig.~\ref{fig4} (c).}
\label{fig7}
\end{figure}

The  LDOS in Fig.~\ref{fig6} are for the Cu/CO system.
As a comparison, if the muffin-tin potential is extremely large (\textit{e.g.} $U_{\textrm{CO}}=$ 9 eV, $d=$ 1.6 nm), it is possible to get a standard band structure of $p_{x,y}$-orbital honeycomb lattice, see Fig.~\ref{fig4} (c-d). We then plot the LDOS of this extreme case  in Fig.~\ref{fig7}. The corresponding band structure is shown in Fig.~\ref{fig4} (c-d), and the energy of the LDOS are denoted by blue arrows.
Fig.~\ref{fig7} (a) is the LDOS of the bottom $p$-orbital flat band. The flat band now is exactly flat, so that the LDOS of this band only shows a bonding state feature. As for the middle two $p$ bands, the LDOS in Fig.~\ref{fig7} (b-c) show bonding and anti-bonding state features, respectively, which are similar to the TB LDOS in Fig.~\ref{fig6} (c-d). The LDOS of the top flat band has a stronger anti-bonding state feature, much obvious than that in Fig.~\ref{fig7} (c).

An interesting case is the new $s$ bands above the $p$ bands in Fig.~\ref{fig4} (c). Though a similar graphene like band structure is found [black lines in Fig.~\ref{fig4} (c)], the new $s$ bands have a different LDOS pattern, as shown in Fig.~\ref{fig7} (d). It is because that the $s$ bands correspond to a new $s$ orbital with a node in the radial part of the wavefunction, which can be obtained by solving Eq.~\eqref{angle} and Eq.~\eqref{radial}.

The LDOS can also be calculated with the DFT simulations. However, due to the finite thickness of Cu slab (8-layer-thick Cu slab) in the DFT simulations, the calculated LDOS contains the contributions from bulk states. It implies that the LDOS calculated with the DFT simulations are not the pure $p$ band LDOS as required. The muffin-tin model should be a better way. And we give the results and discussions about the LDOS from DFT simulations in Appendix~\ref{appb}.

Finally, we argue that the above $p$ band LDOS can also be observed in other similar artificial honeycomb lattice systems, \textit{e.g.}, nano-patterned semiconductor heterostructure~\cite{dvorn2012,polini2013artificial,amo2014,klembt2018}.

\subsection{LDOS pattern: $k \cdot p$ method}\label{sec23}
After diagonalizing the $p_{xy}$-orbital tight-binding Hamiltonian Eq.~\eqref{hamiltonian}, we have coefficients $c^{A}_{p_{x}}(m\textbf{k})$, $c^{A}_{p_{y}}(m\textbf{k})$, $c^{B}_{p_{x}}(m\textbf{k})$ and $c^{B}_{p_{y}}(m\textbf{k})$ for bloch wavefunctions $\{\psi^{A}_{p_x}(\textbf{k},\textbf{r}),\psi^{A}_{p_y}(\textbf{k},\textbf{r}),\psi^{B}_{p_x}(\textbf{k},\textbf{r}),\psi^{B}_{p_y}(\textbf{k},\textbf{r})\}$, which are the basis of TB Hamiltonian. Here, $m$ is the band index. The basis are
\begin{equation}
\psi^{A,B}_{p_{x,y}}(\textbf{k},\textbf{r})=\frac{1}{\sqrt{N}}\sum_{\textbf{R}_{A,B}} \phi^{A,B}_{p_{x,y}}(\textbf{r}-\textbf{R}_{A,B}) e^{i \textbf{k} \textbf{R}_{A,B}},
\end{equation}
where $\phi^{A,B}_{p_{x,y}}(r-R_{A,B})$ are just the wavefunctions of $p$ orbitals given in Eq.~\eqref{porbital}. Then, the eigenfunction of  the $m$-th $p$ band is
\begin{equation}\label{tbldos}
\begin{aligned}
\Psi_{m\textbf{k}}(\textbf{r})=&c^{A}_{p_x}(m\textbf{k})\psi^{A}_{p_x}(\textbf{k},\textbf{r})+c^{A}_{p_y}(m\textbf{k})\psi^{A}_{p_y}(\textbf{k},\textbf{r})\\
+&c^{B}_{p_x}(m\textbf{k})\psi^{B}_{p_x}(\textbf{k},\textbf{r})+c^{B}_{p_y}(m\textbf{k})\psi^{B}_{p_y}(\textbf{k},\textbf{r}).
\end{aligned}
\end{equation}
So, the TB $\textrm{LDOS}$ can be obtained with
\begin{equation}
\textrm{LDOS}(\varepsilon,\textbf{r})=\sum_{m\textbf{k}\sigma}|\Psi_{m\textbf{k}\sigma}(\textbf{r})|^2\delta(\varepsilon-\varepsilon_{m\textbf{k}\sigma}).
\end{equation}

Here, we show how to understand the bonding and antibonding features
of the LDOS shown in Fig.~\ref{fig6} by analyzing the wavefunctions. The wavefunction in Eq.~\eqref{tbldos} can be divided into the components from A site and B site
\begin{equation}
\begin{aligned}
\Psi_{m\textbf{k}}(\textbf{r})=
&\sum_{\textbf{R}_{A}}\frac{e^{i \textbf{k} \textbf{R}_{A}}}{\sqrt{N}}\left\{c^{A}_{p_x}(m\textbf{k})\phi^{A}_{p_x}+c^{A}_{p_y}(m\textbf{k})\phi^{A}_{p_y}\right\}\\
+&\sum_{\textbf{R}_{B}}\frac{e^{i \textbf{k} \textbf{R}_{B}}}{\sqrt{N}}\left\{c^{B}_{p_x}(m\textbf{k})\phi^{B}_{p_x}+c^{B}_{p_y}(m\textbf{k})\phi^{B}_{p_y}\right\},
\end{aligned}
\end{equation}
where the expression in the bracket of the first (second) line is from A (B) site. Near the $\Gamma$ point, the coefficients $c^{A,B}_{p_{x,y}}(m\textbf{k})$ can be got by $k \cdot p$ method~\cite{wu2008}. For example, the wavefunction of the bottom $p$-orbital flat band ($m=1$) near the $\Gamma$ point can be written as
\begin{equation}
\begin{aligned}
\Psi_{1\textbf{k}}(\textbf{r})=
&\sum_{\textbf{R}_{A}}\frac{e^{i \textbf{k} \textbf{R}_{A}}}{\sqrt{2N}|k|}\left\{-(k_y\phi^{A}_{p_x}-k_x\phi^{A}_{p_y})\right\}\\
+&\sum_{\textbf{R}_{B}}\frac{e^{i \textbf{k} \textbf{R}_{B}}}{\sqrt{2N}|k|}\left\{+(k_y\phi^{B}_{p_x}-k_x\phi^{B}_{p_y})\right\}.
\end{aligned}
\end{equation}
We can see that the $p$ orbitals at A and B sites differ in a minus sign. This implies that the two $p$ orbitals are antiparallel (head to head), so that the LDOS in between A and B sites is amplified, i.e., the bonding state feature, as shown in Fig.~\ref{fig6} (c) and (h).

The wavefunctions of other three $p$ bands near the $\Gamma$ point can be written as
\begin{equation}\label{phin2}
\begin{aligned}
\Psi_{2\textbf{k}}(\textbf{r})=
&\sum_{\textbf{R}_{A}}\frac{e^{i \textbf{k} \textbf{R}_{A}}}{\sqrt{2N}|k|}\left\{+(k_x\phi^{A}_{p_x}+k_y\phi^{A}_{p_y})\right\}\\
+&\sum_{\textbf{R}_{B}}\frac{e^{i \textbf{k} \textbf{R}_{B}}}{\sqrt{2N}|k|}\left\{-(k_x\phi^{B}_{p_x}+k_y\phi^{B}_{p_y})\right\},
\end{aligned}
\end{equation}
\begin{equation}\label{phin3}
\begin{aligned}
\Psi_{3\textbf{k}}(\textbf{r})=
&\sum_{\textbf{R}_{A}}\frac{e^{i \textbf{k} \textbf{R}_{A}}}{\sqrt{2N}|k|}\left\{+(k_x\phi^{A}_{p_x}+k_y\phi^{A}_{p_y})\right\}\\
+&\sum_{\textbf{R}_{B}}\frac{e^{i \textbf{k} \textbf{R}_{B}}}{\sqrt{2N}|k|}\left\{+(k_x\phi^{B}_{p_x}+k_y\phi^{B}_{p_y})\right\},
\end{aligned}
\end{equation}
\begin{equation}\label{phin4}
\begin{aligned}
\Psi_{4\textbf{k}}(\textbf{r})=
&\sum_{\textbf{R}_{A}}\frac{e^{i \textbf{k} \textbf{R}_{A}}}{\sqrt{2N}|k|}\left\{-(k_y\phi^{A}_{p_x}-k_x\phi^{A}_{p_y})\right\}\\
+&\sum_{\textbf{R}_{B}}\frac{e^{i \textbf{k} \textbf{R}_{B}}}{\sqrt{2N}|k|}\left\{-(k_y\phi^{B}_{p_x}-k_x\phi^{B}_{p_y})\right\}.
\end{aligned}
\end{equation}
Similarly, the second $p$ band of Eq.~\eqref{phin2} also has a bonding state feature. While for the upper two $p$ bands ($m=3$ and $m=4$), the $p$ orbitals at A site and B site are parallel (head to tail), as shown in Eq.~\eqref{phin3} and Eq.~\eqref{phin4}. Thus the LDOS in between A and B sites is canceled, i.e., the antibonding state feature, as shown in Fig.~\ref{fig6} (d) and (e).

But for LDOS near the K point, the wavefunctions of A and B sites do not have such special relation as that near the $\Gamma$ point~\cite{wu2008}. Thus, we cannot observe a clear bonding or antibonding state feature, as shown in Fig.~\ref{fig6} (b) and (g).

\begin{figure}[ht]
\centering
\includegraphics[width=0.45\textwidth,trim=0 0 0 0,clip]{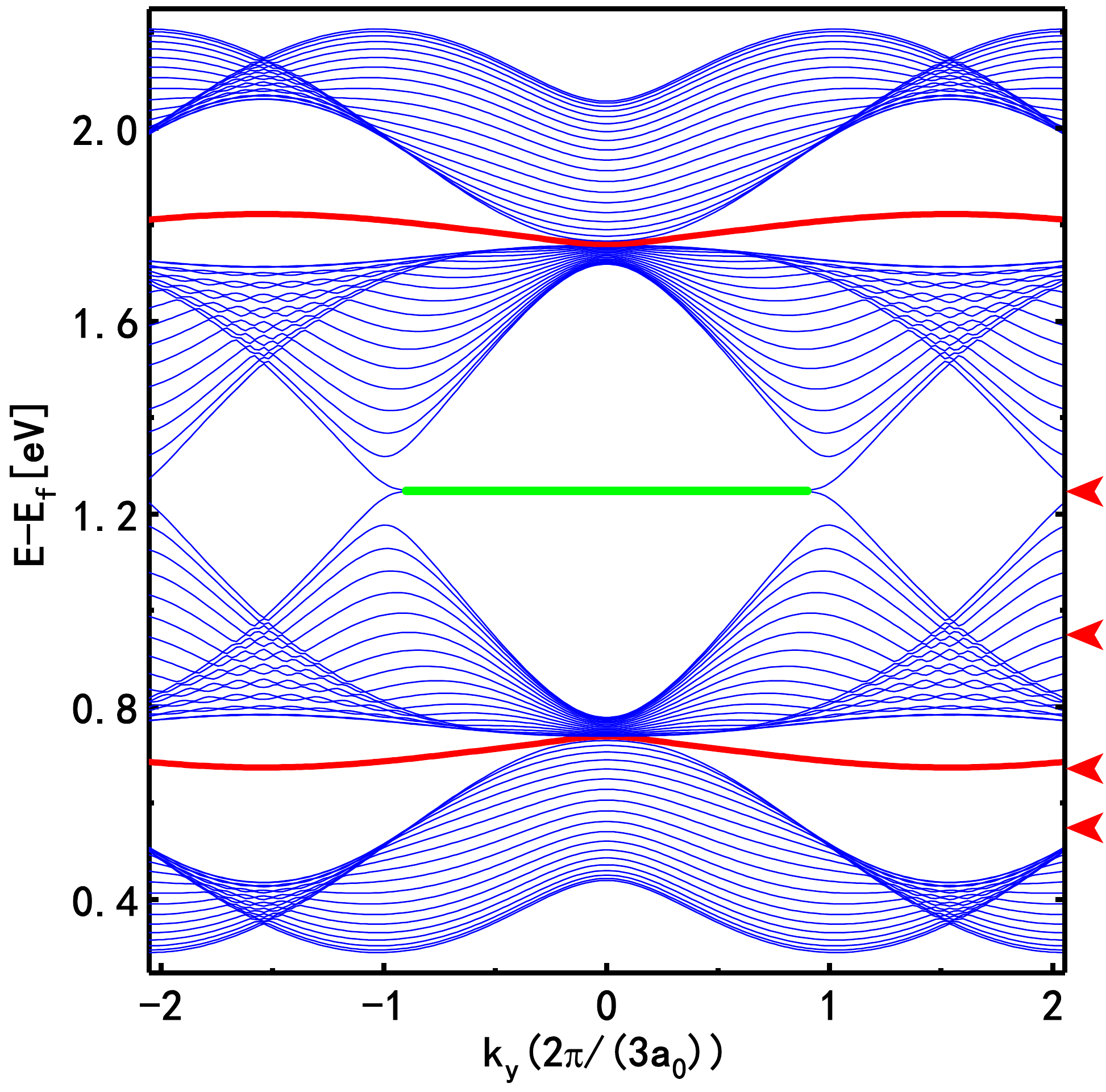}
\caption{(Color online). The $p$-orbital energy bands of a zigzag ribbon of honeycomb lattice. The width of the ribbon is  $29a_0/{\sqrt{3}}$.  We use the  TB model, the parameters and the Fermi level are the same as that of TB bands in Fig.~\ref{fig5} (b). Blue lines are bulk states, green lines are two zero-energy edge states and red lines are two dispersive edge states.}
\label{fig8}
\end{figure}

\subsection{Edge states}
It is pointed out that the  $p_{x,y}$-orbital honeycomb lattice has two kinds of edge states, which have been observed in photonic lattice system~\cite{orbitaledgestate}. Meanwhile, a recent STM experiment has detected the $s$-orbital edge states of the artificial graphene on Cu surface~\cite{wang2014manipulation}.  Thus, it should be ready to measure the edge states of the $p_{x,y}$-orbital honeycomb lattice in this Cu/CO system. We first calculate the bands of a zigzag ribbon of the $p_{x,y}$-orbital honeycomb lattice with the TB model. The width of the zigzag ribbon is about $34$ nm. The calculated bands are given in Fig.~\ref{fig8}. Here, the green lines are the zero energy edge states, and the red lines are the dispersive edge states. These results are all in agreement with former work~\cite{orbitaledgestate}. Note that, in order to compare with Fig.~\ref{fig5} (b), we set $\varepsilon_{p_x}=\varepsilon_{p_y}=1.248$ eV in the TB model for the ribbon calculation.

\begin{figure}[ht]
\centering
\includegraphics[width=0.45\textwidth,trim=0 0 0 0,clip]{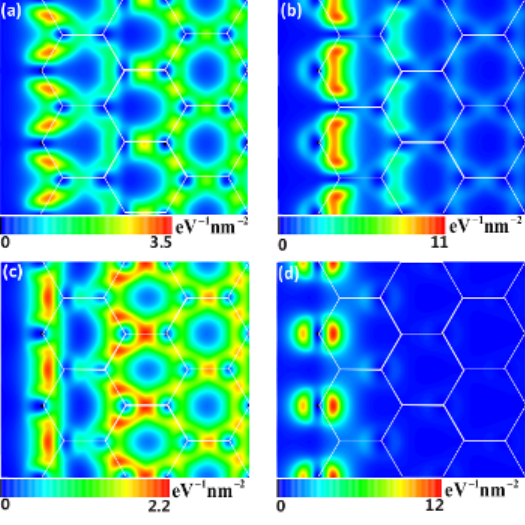}
\caption{(Color online). LDOS patterns near the edge of a zigzag ribbon at  (a) $E-E_f=$ 0.55 eV,  (b) $E-E_f=$ 0.67 eV,  (c) $E-E_f=$ 0.95 eV,  (d) $E-E_f=$ 1.248 eV. Other parameters and the Fermi level are the same as that of ribbon bands in Fig.~\ref{fig8}.}
\label{fig9}
\end{figure}

In Fig.~\ref{fig9}, we plot the LDOS of the zigzag ribbon near a zigzag edge, whose energy are denoted by red arrows in Fig.~\ref{fig8}. Fig.~\ref{fig9} (a) and (c) are the LDOS plotted at $E-E_f=$ 0.55 eV and $E-E_f=$ 0.95 eV, respectively. These LDOS patterns correspond to bulk states, as shown in Fig.~\ref{fig8}. So, the electron density distributes all over the bulk. At $E-E_f=$ 0.67 eV, the states are mainly the dispersive edge states (red lines in Fig.~\ref{fig8}), but there are also some bulk states. So in Fig.~\ref{fig9} (b), we see that the intensity of LDOS is extremely large near the edge, which results from the dispersive edge states. Meanwhile, some bulk states can also be found in Fig.~\ref{fig9} (b). In Fig.~\ref{fig9} (d), we plot the LDOS at $E-E_f=$ 1.248 eV. As shown in Fig.~\ref{fig8}, there are only the zero energy edge states at $E-E_f=$ 1.248 eV. Thus, we see that LDOS becomes nearly zero away from the zigzag edge. Our calculations indicate that we can distinguish different kinds of edge states by measuring the LDOS with STM.

\subsection{Position of Fermi level}
In STM experiment, to access the $p$ orbitals, the Fermi level should be close to $p$ bands. In the Cu/CO system, the position of the Fermi level relative to the surface bands depends on CO-CO (cluster-cluster) spacing $a_0$~\cite{li2016designing,qiu2016designing,ma2017orbital}. In each unit cell of this artificial honeycomb lattice, the number of surface electrons is $S_0 \textrm{N}_e$, where $S_0=\sqrt{3}a_0^2/2$ is the area of unit cell, $\textrm{N}_e \approx  0.72\,\textrm{nm}^{-2}$ is the electron density of Cu surface states. The surface bands can be got by the muffin-tin model. We know that the total number of surface electrons is just the summation of all the surface states below Fermi level. With this relation, we can calculate the position of Fermi level relative to the surface bands, as a function of the CO-CO spacing.

Based on the principle above, we give a simple relation to estimate the required CO-CO spacing, with which one can  let the Fermi level cross the $m$-th surface band. We know that a filled band corresponds to 2 electrons per unit cell (including spin up and spin down). Thus, if the Fermi level crosses the $m$-th surface band, we have $S_0 \textrm{N}_e \leq 2m < S_0 \textrm{N}_e+2$. To let the Fermi level cross the $p$ bands, the relation above give a rough estimation about $a_0$, $2.5\, \textrm{nm}< a_0 <4.4\,\textrm{nm}$, which is our suggested value of the further experiment. Note that, with different value of $a_0$, the surface band shape is similar, and only the band width is changed. This method to predict Fermi level was used in our previous work~\cite{qiu2016designing,ma2017orbital}, and is in good agreement with the experimental observation~\cite{slot2017experimental}.

Here, we point out that the DFT method we used here can not correctly predict the position of the Fermi level, though the modified surface bands can be well achieved. For pure Cu(111) surface, the STM and ARPES experiments indicate that the Fermi level relative to the bottom of  surface band $E_f-E_0$ is about 0.40$\sim$0.45 eV~\cite{kevan1983evidence,jeandupeux1999thermal,tamai2013spin}.
However, DFT simulations with different methods give different values of Fermi level. For example, a full-potential linearized augmented plane-wave (FLAPW) method gives $E_f-E_0=$ 0.526 eV~\cite{butti2005image};
DACAPO code using the ultrasoft pseudopotential plane-wave method gives $E_f-E_0=$ 0.443 eV~\cite{berland2012response};
WIEN97 code using the relativistic self-consistent FLAPW method gives $E_f-E_0=$ 0.40 eV~\cite{courths2001from}; VASP package we used here based on projector-augmented wave method gives $E_f-E_0=$ 0.55 eV [Fig.~\ref{fig3} (a)]. From these studies, we can see that the calculated $E_f-E_0$ has an error range about 150 meV, which is comparable to the width of $s$ bands [about 430 meV, see Fig.~\ref{fig3} (f)].
Thus, we argue that it is quite hard to predict whether the Fermi level crosses the $s$ or $p$ bands in such Cu/CO system by DFT simulations.
Furthermore, when a thin slab is used in the DFT simulations, there is an artificial shift of surface bands (about 150 meV) resulted from the finite size effect, see Fig.~\ref{fig3} (b). It makes the prediction of $E_f$  even worse.

\begin{figure}[t]
\centering
\includegraphics[width=0.45\textwidth,trim=0 0 0 0,clip]{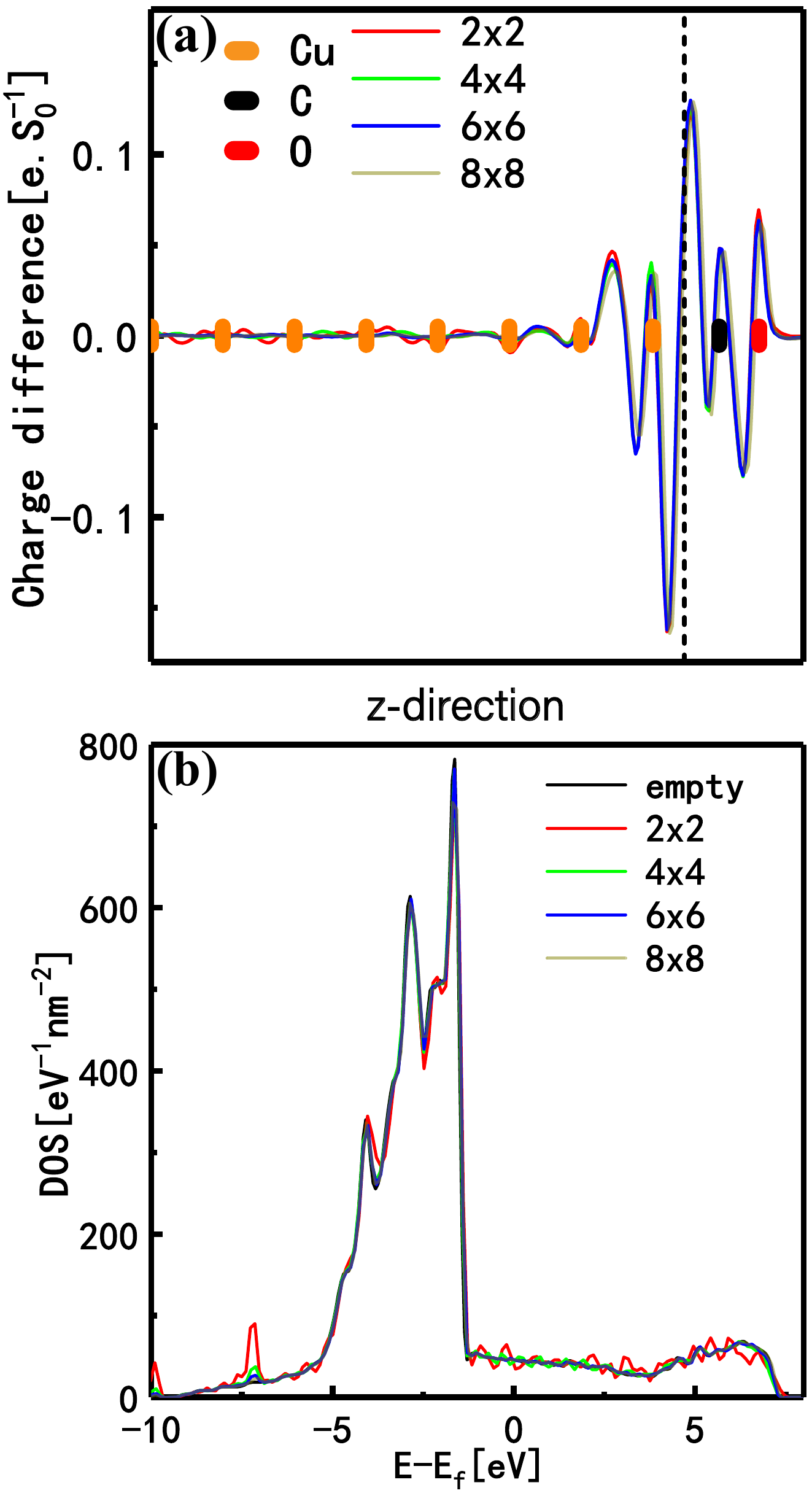}
\caption{(Color online). (a) Charge difference caused by CO adsorption got by DFT simulations with different CO-CO spacing. Other parameters are the same as that used in Fig.~\ref{fig3} (e). The results shown are the values obtained by integrating over the $xy$-plane (parallel to the Cu surface) in one supercell. The horizontal axis represents $z$-direction perpendicular to the Cu surface. Black dashed line represents the Cu/CO interface. (b) The DOS of Cu/CO systems with different CO-CO spacing.}
\label{fig10}
\end{figure}

In above, we assume that the number of Cu (111) surface electrons is fixed even if CO molecules are adsorbed. It is reasonable because that we can not expect one hundred of CO molecules can significantly change the Fermi level of Cu bulk.  To further illustrate this point,
we calculate the CO adsorption induced charge difference per supercell  with different CO spacing. The results are plotted in
Fig.~\ref{fig10} (a). Here, $z$-direction is perpendicular to the Cu surface, and the position of Cu, C and O atoms are denoted in the figure. We see that the charge difference only occurs at the top two Cu surface layers.
The integral of the charge difference along the  $z$-direction gives   a charge transfer of 0.04 $e^{-}$ per supercell from Cu slab to the CO molecules for all the CO spacing used in the calculation. This value is very tiny, considering  the large number of bulk states in  each supercell of the Cu slab. Thus, it indicates that  the charge transfer here can not influence the absolute position of bulk Fermi level. We also calculate the DOS for various cases, with or without CO adatoms, in Fig.~\ref{fig10} (b). There is nearly no change of DOS no matter whether the CO molecules are absorbed or not, and the Fermi level is always at the same position of DOS. All these calculation results suggest that the influence of charge transfer on Fermi level can be safely ignored.

\section{Summary}\label{sec3}
In summary, we illustrate that the desired $p_{x,y}$-orbital honeycomb electron lattice can be realized on Cu surface by depositing CO molecules into a hexagonal lattice with STM tip. By DFT simulations, we get the modified surface bands and the effective muffin-tin parameters, which gives a direct support to the muffin-tin model. Based on these model parameters, we argue that  using a CO cluster instead of a single CO adatom may be a more suitable experimental design to observe the $p$ band features in further experiment. Then, we use the muffin-tin and TB model to illustrate that such Cu/CO systems do give rise to a $p_{x,y}$-orbital honeycomb electron lattice.
We further calculate the LDOS patterns of the $p$ bands  and then give an analytic interpretation with $k \cdot p$ method.
The LDOS of $p$ bands have some unique patterns, which can be used to identify the $p$ bands in further STM experiment.
Finally, we point out that the two kinds of edge states, $\textit{i.e.}$ zero energy and dispersive edge states, can be readily observed in this $p$-orbital honeycomb electron lattice.
We expect that such two kinds of  edge states can be confirmed in further STM experiment.

A further suggestion about experiment is  the lattice constant $a_0$,  namely the distance between adjacent CO atoms.
In order to detect the $p$ bands, the Fermi level should be across the $p$ bands. In the Cu/CO systems, the position of Fermi level depends on $a_0$. Our analysis suggests that an appropriate region is $2.5\, \textrm{nm}< a_0 <4.4\,\textrm{nm}$.

\textit{Note added:}
During the revision of our previous manuscript~\cite{qiu2019making}, we note that Thomas S. Gardenier and collaborators also study the $p$ orbitals in artificial honeycomb lattice in an STM experiment~\cite{gardenier2020}.

\begin{figure*}[t]
\centering
\includegraphics[width=1.0\textwidth,trim=0 0 0 0,clip]{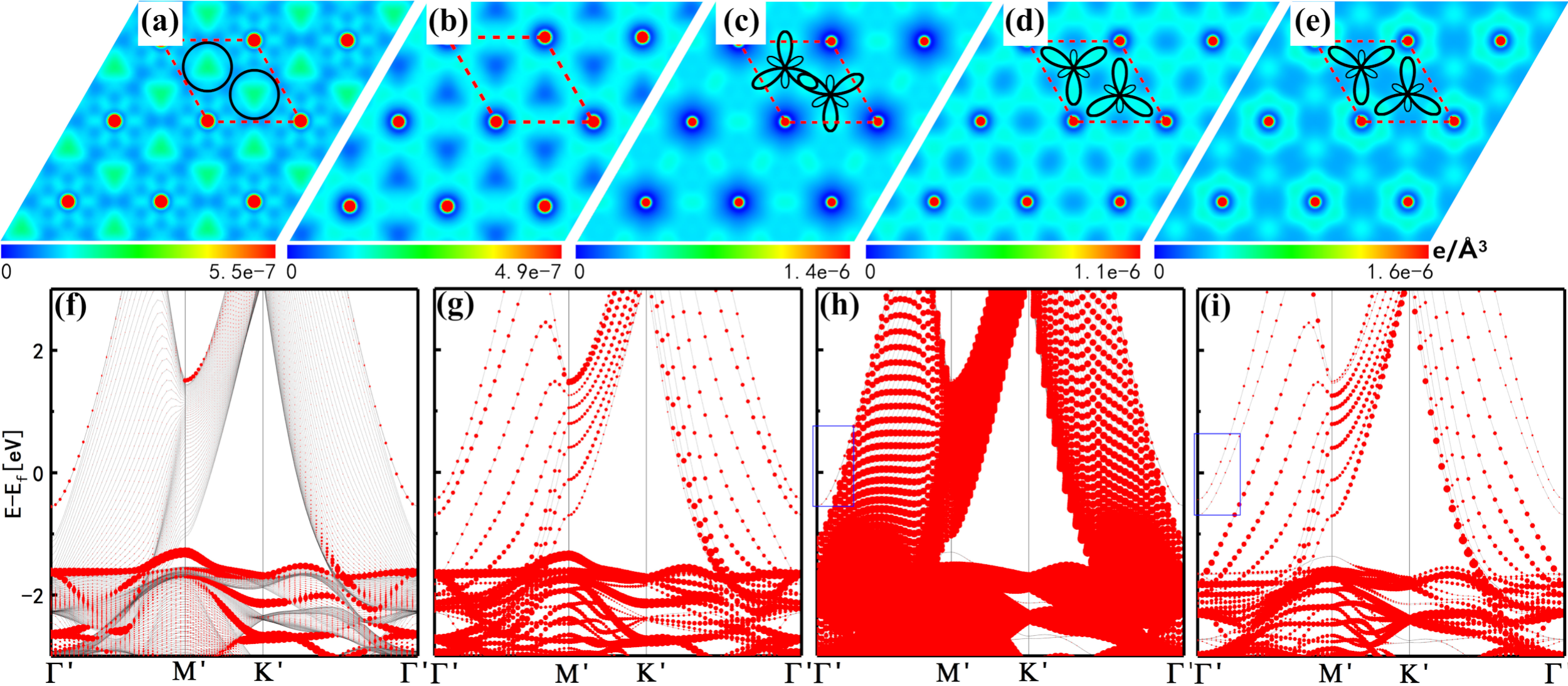}
\caption{(Color online). (a-e) LDOS patterns got by DFT simulations with bias voltage in regions [-0.70 V, -0.69 V], [-0.29 V,-0.28 V], [0.01 V, 0.02 V], [0.19 V, 0.20 V] and [0.33 V, 0.34 V], respectively. In DFT simulations, the Fermi level is set to be zero. The system is the same as that of modified surface bands in Fig.~\ref{fig3} (f). (f-i) Energy bands of a pure Cu slab in a primitive cell got by DFT simulations with projections of surface or medium layers. (f) Bands of a 51-layer slab with projections of surface two layers. (g) Bands of a 8-layer slab with projections of surface two layers. (h) Bands of a 51-layer slab with projections of medium forty seven layers. (i) Bands of a 8-layer slab with projections of medium four layers.}
\label{figs1}
\end{figure*}

\section*{Acknowledgement}
This work was supported by the National Key Research and Development Program of China (Grant No. 2017YFA0403501), the National Natural Science Foundation of China (Grants No. 11874160, 11534001, 21873033, 11274129, 21873033) and the program for HUST academic frontier youth team. Wen-Xuan Qiu also thanks Li-Lan Qin for the helpful support on the DFT simulations.

\appendix

\renewcommand\thefigure{\Alph{section}\arabic{figure}}
\renewcommand\theequation{\Alph{section}\arabic{equation}}

\section{DFT simulations of LDOS}\label{appb}
\setcounter{figure}{0}
\setcounter{equation}{0}

By DFT simulations, we also calculate the LDOS around the Cu surface (Cu/CO system with a $8 \times 8$ supercell). Fig.~\ref{figs1} (a-e) are the LDOS got by DFT simulations with bias voltage in regions [-0.70 V, -0.69 V], [-0.29 V,-0.28 V], [0.01 V, 0.02 V], [0.19 V, 0.20 V] and [0.33 V, 0.34 V], respectively. Note that the energy intervals above all belong to the energy region of the $s$ and $p$ bands in Fig.~\ref{fig3} (f).

Here, we first would like to discuss the relation between the LDOS calculated with the DFT simulations and that with muffin-tin potential model.  The muffin-tin potential model deals with a 2D electron system, so that the unit of LDOS ${\rho}_{sf}^{2D}$ got by Eq.~\eqref{mtldos} is $\textrm{eV}^{-1}\textrm{nm}^{-2}$. As for the DFT simulations, we use a 3D Cu slab to mimic the surface states. The corresponding LDOS is actually the 3D partial charge density at Cu surface in a specified energy interval $\Delta E$, namely, ${\rho}_{sf}^{3D}$, so that the unit is e/$\textrm{\AA}^{3}$. We have a simple relation
\begin{equation}
\lim_{\Delta E\rightarrow 0}\frac{\rho^{3D}_{sf}\times\Delta h}{\Delta E}=\rho^{2D}_{sf},
\end{equation}
where, $\Delta h$ is the thickness of the surface states and is a constant. When we calculate the LDOS at Cu surface with DFT simulations, we choose very small $\Delta E$, in which the 2D surface states dominate if the slab is thick enough, so ${\rho}_{sf}^{3D}\propto{\rho}_{sf}^{2D}$. In this case, the obtained DFT results indeed reflect the LDOS of surface state electrons and are qualitatively in agreement with that from muffin-tin potential model.

In principle, near the surface, the LDOS in the energy region of surface bands  should only result from the Cu surface states.
However, in the DFT simulations, if the thickness of the Cu slab is finite, the contribution of the bulk bands can not be completely ignored.  To illustrate this point, we project all the bands of a pure Cu slab into the surface layers (top and bottom two), in order to see their contributions to the surface LDOS. Note that, in Fig.~\ref{fig3}, we project the bands to a special orbital of top two surface layers so as to identify the surface bands. But here in Fig.~\ref{figs1} we need to project to  all the orbitals of surface layers, since that LDOS contains all the surface orbitals. In Fig.~\ref{figs1} (a), we plot the results of a 51-layer-thick Cu slab, where the surface bands in this case are in Fig.~\ref{fig3} (a). We see that, in the energy region of the surface bands, no other bands contribute to the surface layers, except the surface bands. Fig.~\ref{figs1} (b) gives the same projection for a 8-layer-thick Cu slab, where the corresponding surface bands are plotted in Fig.~\ref{fig3} (b). We see that, except for the surface bands, other bulk bands also have distribution in the surface layers. It implies that the surface LDOS got from the DFT simulations contains some contributions from bulk bands, which is different from the real situation. Therefore, compared with the DFT simulations, the muffin-tin model can give a more reliable description about the $p$ band LDOS patterns of the Cu/CO systems.

Fortunately, despite the interference from the bulk bands, the LDOS got by DFT simulations in Fig.~\ref{figs1} (a-e) are still qualitatively in agreement with that from the muffin-tin model, see in Fig.~\ref{fig6} (f-j). When we increase the energy  from the bottom of surface band, all the typical LDOS of the $p$ bands in Fig.~\ref{fig3} (f) are found in the DFT simulations.

Finally, we show that, even in the 8-layer-thick Cu slab, the surface bands identified in Fig.~\ref{fig3} are indeed localized at the surface layers. Our way is also the layer resolved band projection. Here, we project all the bands of a Cu slab to  the bulk part of the Cu slab. For example, in Fig.~\ref{figs1} (h), we consider the case of a 51-layer-thick Cu slab, where we project the bands into the inner 47 layers.  We see that the surface bands here have nearly no contribution to the bulk layers, see the blue box. It indicates that in this case the surface bands are strongly localized at the Cu surface. The situation is similar in the case of 8-layer-thick Cu slab, as shown in Fig.~\ref{figs1} (i).


\begin{thebibliography}{65}%
\makeatletter
\providecommand \@ifxundefined [1]{%
 \@ifx{#1\undefined}
}%
\providecommand \@ifnum [1]{%
 \ifnum #1\expandafter \@firstoftwo
 \else \expandafter \@secondoftwo
 \fi
}%
\providecommand \@ifx [1]{%
 \ifx #1\expandafter \@firstoftwo
 \else \expandafter \@secondoftwo
 \fi
}%
\providecommand \natexlab [1]{#1}%
\providecommand \enquote  [1]{``#1''}%
\providecommand \bibnamefont  [1]{#1}%
\providecommand \bibfnamefont [1]{#1}%
\providecommand \citenamefont [1]{#1}%
\providecommand \href@noop [0]{\@secondoftwo}%
\providecommand \href [0]{\begingroup \@sanitize@url \@href}%
\providecommand \@href[1]{\@@startlink{#1}\@@href}%
\providecommand \@@href[1]{\endgroup#1\@@endlink}%
\providecommand \@sanitize@url [0]{\catcode `\\12\catcode `\$12\catcode
  `\&12\catcode `\#12\catcode `\^12\catcode `\_12\catcode `\%12\relax}%
\providecommand \@@startlink[1]{}%
\providecommand \@@endlink[0]{}%
\providecommand \url  [0]{\begingroup\@sanitize@url \@url }%
\providecommand \@url [1]{\endgroup\@href {#1}{\urlprefix }}%
\providecommand \urlprefix  [0]{URL }%
\providecommand \Eprint [0]{\href }%
\providecommand \doibase [0]{http://dx.doi.org/}%
\providecommand \selectlanguage [0]{\@gobble}%
\providecommand \bibinfo  [0]{\@secondoftwo}%
\providecommand \bibfield  [0]{\@secondoftwo}%
\providecommand \translation [1]{[#1]}%
\providecommand \BibitemOpen [0]{}%
\providecommand \bibitemStop [0]{}%
\providecommand \bibitemNoStop [0]{.\EOS\space}%
\providecommand \EOS [0]{\spacefactor3000\relax}%
\providecommand \BibitemShut  [1]{\csname bibitem#1\endcsname}%
\let\auto@bib@innerbib\@empty
\bibitem [{\citenamefont {Millis}\ \emph {et~al.}(1996)\citenamefont {Millis},
  \citenamefont {Shraiman},\ and\ \citenamefont
  {Mueller}}]{PhysRevLett.77.175}%
  \BibitemOpen
  \bibfield  {author} {\bibinfo {author} {\bibfnamefont {A.~J.}\ \bibnamefont
  {Millis}}, \bibinfo {author} {\bibfnamefont {B.~I.}\ \bibnamefont
  {Shraiman}}, \ and\ \bibinfo {author} {\bibfnamefont {R.}~\bibnamefont
  {Mueller}},\ }\href {\doibase 10.1103/PhysRevLett.77.175} {\bibfield
  {journal} {\bibinfo  {journal} {Phys. Rev. Lett.}\ }\textbf {\bibinfo
  {volume} {77}},\ \bibinfo {pages} {175} (\bibinfo {year} {1996})}\BibitemShut
  {NoStop}%
\bibitem [{\citenamefont {Ramirez}(1997)}]{ramirez1997colossal}%
  \BibitemOpen
  \bibfield  {author} {\bibinfo {author} {\bibfnamefont {A.}~\bibnamefont
  {Ramirez}},\ }\href {https://doi.org/10.1088/0953-8984/9/39/005} {\bibfield
  {journal} {\bibinfo  {journal} {J. Phys.: Condens. Matter}\ }\textbf
  {\bibinfo {volume} {9}},\ \bibinfo {pages} {8171} (\bibinfo {year}
  {1997})}\BibitemShut {NoStop}%
\bibitem [{\citenamefont {Ling}\ \emph {et~al.}(2000)\citenamefont {Ling},
  \citenamefont {Millburn}, \citenamefont {Mitchell}, \citenamefont {Argyriou},
  \citenamefont {Linton},\ and\ \citenamefont {Bordallo}}]{PhysRevB.62.15096}%
  \BibitemOpen
  \bibfield  {author} {\bibinfo {author} {\bibfnamefont {C.~D.}\ \bibnamefont
  {Ling}}, \bibinfo {author} {\bibfnamefont {J.~E.}\ \bibnamefont {Millburn}},
  \bibinfo {author} {\bibfnamefont {J.~F.}\ \bibnamefont {Mitchell}}, \bibinfo
  {author} {\bibfnamefont {D.~N.}\ \bibnamefont {Argyriou}}, \bibinfo {author}
  {\bibfnamefont {J.}~\bibnamefont {Linton}}, \ and\ \bibinfo {author}
  {\bibfnamefont {H.~N.}\ \bibnamefont {Bordallo}},\ }\href {\doibase
  10.1103/PhysRevB.62.15096} {\bibfield  {journal} {\bibinfo  {journal} {Phys.
  Rev. B}\ }\textbf {\bibinfo {volume} {62}},\ \bibinfo {pages} {15096}
  (\bibinfo {year} {2000})}\BibitemShut {NoStop}%
\bibitem [{\citenamefont {Mathieu}\ \emph {et~al.}(2004)\citenamefont
  {Mathieu}, \citenamefont {Akahoshi}, \citenamefont {Asamitsu}, \citenamefont
  {Tomioka},\ and\ \citenamefont {Tokura}}]{PhysRevLett.93.227202}%
  \BibitemOpen
  \bibfield  {author} {\bibinfo {author} {\bibfnamefont {R.}~\bibnamefont
  {Mathieu}}, \bibinfo {author} {\bibfnamefont {D.}~\bibnamefont {Akahoshi}},
  \bibinfo {author} {\bibfnamefont {A.}~\bibnamefont {Asamitsu}}, \bibinfo
  {author} {\bibfnamefont {Y.}~\bibnamefont {Tomioka}}, \ and\ \bibinfo
  {author} {\bibfnamefont {Y.}~\bibnamefont {Tokura}},\ }\href {\doibase
  10.1103/PhysRevLett.93.227202} {\bibfield  {journal} {\bibinfo  {journal}
  {Phys. Rev. Lett.}\ }\textbf {\bibinfo {volume} {93}},\ \bibinfo {pages}
  {227202} (\bibinfo {year} {2004})}\BibitemShut {NoStop}%
\bibitem [{\citenamefont {Bednorz}\ and\ \citenamefont
  {M{\"u}ller}(1986)}]{Bednorz1986}%
  \BibitemOpen
  \bibfield  {author} {\bibinfo {author} {\bibfnamefont {J.~G.}\ \bibnamefont
  {Bednorz}}\ and\ \bibinfo {author} {\bibfnamefont {K.~A.}\ \bibnamefont
  {M{\"u}ller}},\ }\href {\doibase 10.1007/BF01303701} {\bibfield  {journal}
  {\bibinfo  {journal} {Z. Phys. B}\ }\textbf {\bibinfo {volume} {64}},\
  \bibinfo {pages} {189} (\bibinfo {year} {1986})}\BibitemShut {NoStop}%
\bibitem [{\citenamefont {Luke}\ \emph {et~al.}(1998)\citenamefont {Luke},
  \citenamefont {Fudamoto}, \citenamefont {Kojima}, \citenamefont {Larkin},
  \citenamefont {Merrin}, \citenamefont {Nachumi}, \citenamefont {Uemura},
  \citenamefont {Maeno}, \citenamefont {Mao}, \citenamefont {Mori} \emph
  {et~al.}}]{luke1998time}%
  \BibitemOpen
  \bibfield  {author} {\bibinfo {author} {\bibfnamefont {G.~M.}\ \bibnamefont
  {Luke}}, \bibinfo {author} {\bibfnamefont {Y.}~\bibnamefont {Fudamoto}},
  \bibinfo {author} {\bibfnamefont {K.}~\bibnamefont {Kojima}}, \bibinfo
  {author} {\bibfnamefont {M.}~\bibnamefont {Larkin}}, \bibinfo {author}
  {\bibfnamefont {J.}~\bibnamefont {Merrin}}, \bibinfo {author} {\bibfnamefont
  {B.}~\bibnamefont {Nachumi}}, \bibinfo {author} {\bibfnamefont
  {Y.}~\bibnamefont {Uemura}}, \bibinfo {author} {\bibfnamefont
  {Y.}~\bibnamefont {Maeno}}, \bibinfo {author} {\bibfnamefont
  {Z.}~\bibnamefont {Mao}}, \bibinfo {author} {\bibfnamefont {Y.}~\bibnamefont
  {Mori}},  \emph {et~al.},\ }\href {https://doi.org/10.1038/29038} {\bibfield
  {journal} {\bibinfo  {journal} {Nature}\ }\textbf {\bibinfo {volume} {394}},\
  \bibinfo {pages} {558} (\bibinfo {year} {1998})}\BibitemShut {NoStop}%
\bibitem [{\citenamefont {Ohtomo}\ and\ \citenamefont
  {Hwang}(2004)}]{ohtomo2004high}%
  \BibitemOpen
  \bibfield  {author} {\bibinfo {author} {\bibfnamefont {A.}~\bibnamefont
  {Ohtomo}}\ and\ \bibinfo {author} {\bibfnamefont {H.}~\bibnamefont {Hwang}},\
  }\href {https://doi.org/10.1038/nature02308} {\bibfield  {journal} {\bibinfo
  {journal} {Nature}\ }\textbf {\bibinfo {volume} {427}},\ \bibinfo {pages}
  {423} (\bibinfo {year} {2004})}\BibitemShut {NoStop}%
\bibitem [{\citenamefont {Kamihara}\ \emph {et~al.}(2006)\citenamefont
  {Kamihara}, \citenamefont {Hiramatsu}, \citenamefont {Hirano}, \citenamefont
  {Kawamura}, \citenamefont {Yanagi}, \citenamefont {Kamiya},\ and\
  \citenamefont {Hosono}}]{ja063355c}%
  \BibitemOpen
  \bibfield  {author} {\bibinfo {author} {\bibfnamefont {Y.}~\bibnamefont
  {Kamihara}}, \bibinfo {author} {\bibfnamefont {H.}~\bibnamefont {Hiramatsu}},
  \bibinfo {author} {\bibfnamefont {M.}~\bibnamefont {Hirano}}, \bibinfo
  {author} {\bibfnamefont {R.}~\bibnamefont {Kawamura}}, \bibinfo {author}
  {\bibfnamefont {H.}~\bibnamefont {Yanagi}}, \bibinfo {author} {\bibfnamefont
  {T.}~\bibnamefont {Kamiya}}, \ and\ \bibinfo {author} {\bibfnamefont
  {H.}~\bibnamefont {Hosono}},\ }\href {\doibase 10.1021/ja063355c} {\bibfield
  {journal} {\bibinfo  {journal} {J. Am. Chem. Soc.}\ }\textbf {\bibinfo
  {volume} {128}},\ \bibinfo {pages} {10012} (\bibinfo {year}
  {2006})}\BibitemShut {NoStop}%
\bibitem [{\citenamefont {Van~Aken}\ \emph {et~al.}(2003)\citenamefont
  {Van~Aken}, \citenamefont {Jurchescu}, \citenamefont {Meetsma}, \citenamefont
  {Tomioka}, \citenamefont {Tokura},\ and\ \citenamefont
  {Palstra}}]{PhysRevLett.90.066403}%
  \BibitemOpen
  \bibfield  {author} {\bibinfo {author} {\bibfnamefont {B.~B.}\ \bibnamefont
  {Van~Aken}}, \bibinfo {author} {\bibfnamefont {O.~D.}\ \bibnamefont
  {Jurchescu}}, \bibinfo {author} {\bibfnamefont {A.}~\bibnamefont {Meetsma}},
  \bibinfo {author} {\bibfnamefont {Y.}~\bibnamefont {Tomioka}}, \bibinfo
  {author} {\bibfnamefont {Y.}~\bibnamefont {Tokura}}, \ and\ \bibinfo {author}
  {\bibfnamefont {T.~T.~M.}\ \bibnamefont {Palstra}},\ }\href {\doibase
  10.1103/PhysRevLett.90.066403} {\bibfield  {journal} {\bibinfo  {journal}
  {Phys. Rev. Lett.}\ }\textbf {\bibinfo {volume} {90}},\ \bibinfo {pages}
  {066403} (\bibinfo {year} {2003})}\BibitemShut {NoStop}%
\bibitem [{\citenamefont {Chen}\ and\ \citenamefont
  {Zou}(2007)}]{chen2007orbital}%
  \BibitemOpen
  \bibfield  {author} {\bibinfo {author} {\bibfnamefont {D.-M.}\ \bibnamefont
  {Chen}}\ and\ \bibinfo {author} {\bibfnamefont {L.-J.}\ \bibnamefont {Zou}},\
  }\href {https://doi.org/10.1142/S0217979207036618} {\bibfield  {journal}
  {\bibinfo  {journal} {Int. J. Mod. Phys. B}\ }\textbf {\bibinfo {volume}
  {21}},\ \bibinfo {pages} {691} (\bibinfo {year} {2007})}\BibitemShut
  {NoStop}%
\bibitem [{\citenamefont {Kong}\ \emph {et~al.}(2008)\citenamefont {Kong},
  \citenamefont {Zhang},\ and\ \citenamefont {Shi}}]{kong2008orbital}%
  \BibitemOpen
  \bibfield  {author} {\bibinfo {author} {\bibfnamefont {S.}~\bibnamefont
  {Kong}}, \bibinfo {author} {\bibfnamefont {W.}~\bibnamefont {Zhang}}, \ and\
  \bibinfo {author} {\bibfnamefont {D.}~\bibnamefont {Shi}},\ }\href
  {https://doi.org/10.1088/1367-2630/10/9/093020} {\bibfield  {journal}
  {\bibinfo  {journal} {New J. Phys.}\ }\textbf {\bibinfo {volume} {10}},\
  \bibinfo {pages} {093020} (\bibinfo {year} {2008})}\BibitemShut {NoStop}%
\bibitem [{\citenamefont {M\"uller}\ \emph {et~al.}(2007)\citenamefont
  {M\"uller}, \citenamefont {F\"olling}, \citenamefont {Widera},\ and\
  \citenamefont {Bloch}}]{PhysRevLett.99.200405}%
  \BibitemOpen
  \bibfield  {author} {\bibinfo {author} {\bibfnamefont {T.}~\bibnamefont
  {M\"uller}}, \bibinfo {author} {\bibfnamefont {S.}~\bibnamefont {F\"olling}},
  \bibinfo {author} {\bibfnamefont {A.}~\bibnamefont {Widera}}, \ and\ \bibinfo
  {author} {\bibfnamefont {I.}~\bibnamefont {Bloch}},\ }\href {\doibase
  10.1103/PhysRevLett.99.200405} {\bibfield  {journal} {\bibinfo  {journal}
  {Phys. Rev. Lett.}\ }\textbf {\bibinfo {volume} {99}},\ \bibinfo {pages}
  {200405} (\bibinfo {year} {2007})}\BibitemShut {NoStop}%
\bibitem [{\citenamefont {Mili\ifmmode \acute{c}\else
  \'{c}\fi{}evi\ifmmode~\acute{c}\else \'{c}\fi{}}\ \emph
  {et~al.}(2017)\citenamefont {Mili\ifmmode \acute{c}\else
  \'{c}\fi{}evi\ifmmode~\acute{c}\else \'{c}\fi{}}, \citenamefont {Ozawa},
  \citenamefont {Montambaux}, \citenamefont {Carusotto}, \citenamefont
  {Galopin}, \citenamefont {Lema\^{\i}tre}, \citenamefont {Le~Gratiet},
  \citenamefont {Sagnes}, \citenamefont {Bloch},\ and\ \citenamefont
  {Amo}}]{orbitaledgestate}%
  \BibitemOpen
  \bibfield  {author} {\bibinfo {author} {\bibfnamefont {M.}~\bibnamefont
  {Mili\ifmmode \acute{c}\else \'{c}\fi{}evi\ifmmode~\acute{c}\else
  \'{c}\fi{}}}, \bibinfo {author} {\bibfnamefont {T.}~\bibnamefont {Ozawa}},
  \bibinfo {author} {\bibfnamefont {G.}~\bibnamefont {Montambaux}}, \bibinfo
  {author} {\bibfnamefont {I.}~\bibnamefont {Carusotto}}, \bibinfo {author}
  {\bibfnamefont {E.}~\bibnamefont {Galopin}}, \bibinfo {author} {\bibfnamefont
  {A.}~\bibnamefont {Lema\^{\i}tre}}, \bibinfo {author} {\bibfnamefont
  {L.}~\bibnamefont {Le~Gratiet}}, \bibinfo {author} {\bibfnamefont
  {I.}~\bibnamefont {Sagnes}}, \bibinfo {author} {\bibfnamefont
  {J.}~\bibnamefont {Bloch}}, \ and\ \bibinfo {author} {\bibfnamefont
  {A.}~\bibnamefont {Amo}},\ }\href {\doibase 10.1103/PhysRevLett.118.107403}
  {\bibfield  {journal} {\bibinfo  {journal} {Phys. Rev. Lett.}\ }\textbf
  {\bibinfo {volume} {118}},\ \bibinfo {pages} {107403} (\bibinfo {year}
  {2017})}\BibitemShut {NoStop}%
\bibitem [{\citenamefont {Wu}(2009)}]{wu2009unconventional}%
  \BibitemOpen
  \bibfield  {author} {\bibinfo {author} {\bibfnamefont {C.}~\bibnamefont
  {Wu}},\ }\href {https://doi.org/10.1142/S0217984909017777} {\bibfield
  {journal} {\bibinfo  {journal} {Mod. Phys. Lett.}\ }\textbf {\bibinfo
  {volume} {23}},\ \bibinfo {pages} {1} (\bibinfo {year} {2009})}\BibitemShut
  {NoStop}%
\bibitem [{\citenamefont {Li}\ and\ \citenamefont {Liu}(2016)}]{li2016physics}%
  \BibitemOpen
  \bibfield  {author} {\bibinfo {author} {\bibfnamefont {X.}~\bibnamefont
  {Li}}\ and\ \bibinfo {author} {\bibfnamefont {W.~V.}\ \bibnamefont {Liu}},\
  }\href {https://doi.org/10.1088/0034-4885/79/11/116401} {\bibfield  {journal}
  {\bibinfo  {journal} {Rep. Prog. Phys.}\ }\textbf {\bibinfo {volume} {79}},\
  \bibinfo {pages} {116401} (\bibinfo {year} {2016})}\BibitemShut {NoStop}%
\bibitem [{\citenamefont {Wirth}\ \emph {et~al.}(2011)\citenamefont {Wirth},
  \citenamefont {{\"O}lschl{\"a}ger},\ and\ \citenamefont
  {Hemmerich}}]{wirth2011evidence}%
  \BibitemOpen
  \bibfield  {author} {\bibinfo {author} {\bibfnamefont {G.}~\bibnamefont
  {Wirth}}, \bibinfo {author} {\bibfnamefont {M.}~\bibnamefont
  {{\"O}lschl{\"a}ger}}, \ and\ \bibinfo {author} {\bibfnamefont
  {A.}~\bibnamefont {Hemmerich}},\ }\href {https://doi.org/10.1038/nphys1857}
  {\bibfield  {journal} {\bibinfo  {journal} {Nat. Phys.}\ }\textbf {\bibinfo
  {volume} {7}},\ \bibinfo {pages} {147} (\bibinfo {year} {2011})}\BibitemShut
  {NoStop}%
\bibitem [{\citenamefont {{\"O}lschl{\"a}ger}\ \emph
  {et~al.}(2013)\citenamefont {{\"O}lschl{\"a}ger}, \citenamefont {Kock},
  \citenamefont {Wirth}, \citenamefont {Ewerbeck}, \citenamefont {Smith},\ and\
  \citenamefont {Hemmerich}}]{olschlager2013interaction}%
  \BibitemOpen
  \bibfield  {author} {\bibinfo {author} {\bibfnamefont {M.}~\bibnamefont
  {{\"O}lschl{\"a}ger}}, \bibinfo {author} {\bibfnamefont {T.}~\bibnamefont
  {Kock}}, \bibinfo {author} {\bibfnamefont {G.}~\bibnamefont {Wirth}},
  \bibinfo {author} {\bibfnamefont {A.}~\bibnamefont {Ewerbeck}}, \bibinfo
  {author} {\bibfnamefont {C.~M.}\ \bibnamefont {Smith}}, \ and\ \bibinfo
  {author} {\bibfnamefont {A.}~\bibnamefont {Hemmerich}},\ }\href
  {https://doi.org/10.1088/1367-2630/15/8/083041} {\bibfield  {journal}
  {\bibinfo  {journal} {New J. Phys.}\ }\textbf {\bibinfo {volume} {15}},\
  \bibinfo {pages} {083041} (\bibinfo {year} {2013})}\BibitemShut {NoStop}%
\bibitem [{\citenamefont {Kock}\ \emph {et~al.}(2015)\citenamefont {Kock},
  \citenamefont {\"Olschl\"ager}, \citenamefont {Ewerbeck}, \citenamefont
  {Huang}, \citenamefont {Mathey},\ and\ \citenamefont
  {Hemmerich}}]{PhysRevLett.114.115301}%
  \BibitemOpen
  \bibfield  {author} {\bibinfo {author} {\bibfnamefont {T.}~\bibnamefont
  {Kock}}, \bibinfo {author} {\bibfnamefont {M.}~\bibnamefont
  {\"Olschl\"ager}}, \bibinfo {author} {\bibfnamefont {A.}~\bibnamefont
  {Ewerbeck}}, \bibinfo {author} {\bibfnamefont {W.-M.}\ \bibnamefont {Huang}},
  \bibinfo {author} {\bibfnamefont {L.}~\bibnamefont {Mathey}}, \ and\ \bibinfo
  {author} {\bibfnamefont {A.}~\bibnamefont {Hemmerich}},\ }\href {\doibase
  10.1103/PhysRevLett.114.115301} {\bibfield  {journal} {\bibinfo  {journal}
  {Phys. Rev. Lett.}\ }\textbf {\bibinfo {volume} {114}},\ \bibinfo {pages}
  {115301} (\bibinfo {year} {2015})}\BibitemShut {NoStop}%
\bibitem [{\citenamefont {Wu}\ \emph {et~al.}(2007)\citenamefont {Wu},
  \citenamefont {Bergman}, \citenamefont {Balents},\ and\ \citenamefont
  {Das~Sarma}}]{wu2007}%
  \BibitemOpen
  \bibfield  {author} {\bibinfo {author} {\bibfnamefont {C.}~\bibnamefont
  {Wu}}, \bibinfo {author} {\bibfnamefont {D.}~\bibnamefont {Bergman}},
  \bibinfo {author} {\bibfnamefont {L.}~\bibnamefont {Balents}}, \ and\
  \bibinfo {author} {\bibfnamefont {S.}~\bibnamefont {Das~Sarma}},\ }\href
  {\doibase 10.1103/PhysRevLett.99.070401} {\bibfield  {journal} {\bibinfo
  {journal} {Phys. Rev. Lett.}\ }\textbf {\bibinfo {volume} {99}},\ \bibinfo
  {pages} {070401} (\bibinfo {year} {2007})}\BibitemShut {NoStop}%
\bibitem [{\citenamefont {Wu}\ and\ \citenamefont {Das~Sarma}(2008)}]{wu2008}%
  \BibitemOpen
  \bibfield  {author} {\bibinfo {author} {\bibfnamefont {C.}~\bibnamefont
  {Wu}}\ and\ \bibinfo {author} {\bibfnamefont {S.}~\bibnamefont {Das~Sarma}},\
  }\href {\doibase 10.1103/PhysRevB.77.235107} {\bibfield  {journal} {\bibinfo
  {journal} {Phys. Rev. B}\ }\textbf {\bibinfo {volume} {77}},\ \bibinfo
  {pages} {235107} (\bibinfo {year} {2008})}\BibitemShut {NoStop}%
\bibitem [{\citenamefont {Zhang}\ \emph {et~al.}(2010)\citenamefont {Zhang},
  \citenamefont {Hung},\ and\ \citenamefont {Wu}}]{PhysRevA.82.053618}%
  \BibitemOpen
  \bibfield  {author} {\bibinfo {author} {\bibfnamefont {S.}~\bibnamefont
  {Zhang}}, \bibinfo {author} {\bibfnamefont {H.-h.}\ \bibnamefont {Hung}}, \
  and\ \bibinfo {author} {\bibfnamefont {C.}~\bibnamefont {Wu}},\ }\href
  {\doibase 10.1103/PhysRevA.82.053618} {\bibfield  {journal} {\bibinfo
  {journal} {Phys. Rev. A}\ }\textbf {\bibinfo {volume} {82}},\ \bibinfo
  {pages} {053618} (\bibinfo {year} {2010})}\BibitemShut {NoStop}%
\bibitem [{\citenamefont {Zhang}\ \emph {et~al.}(2019)\citenamefont {Zhang},
  \citenamefont {Wang},\ and\ \citenamefont {Zhang}}]{Zhang2019}%
  \BibitemOpen
  \bibfield  {author} {\bibinfo {author} {\bibfnamefont {C.}~\bibnamefont
  {Zhang}}, \bibinfo {author} {\bibfnamefont {Y.}~\bibnamefont {Wang}}, \ and\
  \bibinfo {author} {\bibfnamefont {W.}~\bibnamefont {Zhang}},\ }\href
  {\doibase 10.1088/1361-648x/ab2289} {\bibfield  {journal} {\bibinfo
  {journal} {J. Phys.: Condens. Matter}\ }\textbf {\bibinfo {volume} {31}},\
  \bibinfo {pages} {335403} (\bibinfo {year} {2019})}\BibitemShut {NoStop}%
\bibitem [{\citenamefont {Zhu}\ \emph {et~al.}(2019)\citenamefont {Zhu},
  \citenamefont {Sun}, \citenamefont {Yang}, \citenamefont {Wang},
  \citenamefont {Liu},\ and\ \citenamefont {Ji}}]{ZhuGB2019}%
  \BibitemOpen
  \bibfield  {author} {\bibinfo {author} {\bibfnamefont {G.-B.}\ \bibnamefont
  {Zhu}}, \bibinfo {author} {\bibfnamefont {Q.}~\bibnamefont {Sun}}, \bibinfo
  {author} {\bibfnamefont {H.-M.}\ \bibnamefont {Yang}}, \bibinfo {author}
  {\bibfnamefont {L.-L.}\ \bibnamefont {Wang}}, \bibinfo {author}
  {\bibfnamefont {W.-M.}\ \bibnamefont {Liu}}, \ and\ \bibinfo {author}
  {\bibfnamefont {A.-C.}\ \bibnamefont {Ji}},\ }\href {\doibase
  10.1103/PhysRevA.100.043608} {\bibfield  {journal} {\bibinfo  {journal}
  {Phys. Rev. A}\ }\textbf {\bibinfo {volume} {100}},\ \bibinfo {pages}
  {043608} (\bibinfo {year} {2019})}\BibitemShut {NoStop}%
\bibitem [{\citenamefont {Wu}(2008)}]{PhysRevLett.101.186807}%
  \BibitemOpen
  \bibfield  {author} {\bibinfo {author} {\bibfnamefont {C.}~\bibnamefont
  {Wu}},\ }\href {\doibase 10.1103/PhysRevLett.101.186807} {\bibfield
  {journal} {\bibinfo  {journal} {Phys. Rev. Lett.}\ }\textbf {\bibinfo
  {volume} {101}},\ \bibinfo {pages} {186807} (\bibinfo {year}
  {2008})}\BibitemShut {NoStop}%
\bibitem [{\citenamefont {Zhang}\ \emph {et~al.}(2011)\citenamefont {Zhang},
  \citenamefont {Hung}, \citenamefont {Zhang},\ and\ \citenamefont
  {Wu}}]{PhysRevA.83.023615}%
  \BibitemOpen
  \bibfield  {author} {\bibinfo {author} {\bibfnamefont {M.}~\bibnamefont
  {Zhang}}, \bibinfo {author} {\bibfnamefont {H.-h.}\ \bibnamefont {Hung}},
  \bibinfo {author} {\bibfnamefont {C.}~\bibnamefont {Zhang}}, \ and\ \bibinfo
  {author} {\bibfnamefont {C.}~\bibnamefont {Wu}},\ }\href {\doibase
  10.1103/PhysRevA.83.023615} {\bibfield  {journal} {\bibinfo  {journal} {Phys.
  Rev. A}\ }\textbf {\bibinfo {volume} {83}},\ \bibinfo {pages} {023615}
  (\bibinfo {year} {2011})}\BibitemShut {NoStop}%
\bibitem [{\citenamefont {Zhang}\ \emph {et~al.}(2014)\citenamefont {Zhang},
  \citenamefont {Li},\ and\ \citenamefont {Wu}}]{PhysRevB.90.075114}%
  \BibitemOpen
  \bibfield  {author} {\bibinfo {author} {\bibfnamefont {G.-F.}\ \bibnamefont
  {Zhang}}, \bibinfo {author} {\bibfnamefont {Y.}~\bibnamefont {Li}}, \ and\
  \bibinfo {author} {\bibfnamefont {C.}~\bibnamefont {Wu}},\ }\href {\doibase
  10.1103/PhysRevB.90.075114} {\bibfield  {journal} {\bibinfo  {journal} {Phys.
  Rev. B}\ }\textbf {\bibinfo {volume} {90}},\ \bibinfo {pages} {075114}
  (\bibinfo {year} {2014})}\BibitemShut {NoStop}%
\bibitem [{\citenamefont {Mili\ifmmode \acute{c}\else
  \'{c}\fi{}evi\ifmmode~\acute{c}\else \'{c}\fi{}}\ \emph
  {et~al.}(2019)\citenamefont {Mili\ifmmode \acute{c}\else
  \'{c}\fi{}evi\ifmmode~\acute{c}\else \'{c}\fi{}}, \citenamefont {Montambaux},
  \citenamefont {Ozawa}, \citenamefont {Jamadi}, \citenamefont {Real},
  \citenamefont {Sagnes}, \citenamefont {Lema\^{\i}tre}, \citenamefont
  {Le~Gratiet}, \citenamefont {Harouri}, \citenamefont {Bloch},\ and\
  \citenamefont {Amo}}]{Bloch2019}%
  \BibitemOpen
  \bibfield  {author} {\bibinfo {author} {\bibfnamefont {M.}~\bibnamefont
  {Mili\ifmmode \acute{c}\else \'{c}\fi{}evi\ifmmode~\acute{c}\else
  \'{c}\fi{}}}, \bibinfo {author} {\bibfnamefont {G.}~\bibnamefont
  {Montambaux}}, \bibinfo {author} {\bibfnamefont {T.}~\bibnamefont {Ozawa}},
  \bibinfo {author} {\bibfnamefont {O.}~\bibnamefont {Jamadi}}, \bibinfo
  {author} {\bibfnamefont {B.}~\bibnamefont {Real}}, \bibinfo {author}
  {\bibfnamefont {I.}~\bibnamefont {Sagnes}}, \bibinfo {author} {\bibfnamefont
  {A.}~\bibnamefont {Lema\^{\i}tre}}, \bibinfo {author} {\bibfnamefont
  {L.}~\bibnamefont {Le~Gratiet}}, \bibinfo {author} {\bibfnamefont
  {A.}~\bibnamefont {Harouri}}, \bibinfo {author} {\bibfnamefont
  {J.}~\bibnamefont {Bloch}}, \ and\ \bibinfo {author} {\bibfnamefont
  {A.}~\bibnamefont {Amo}},\ }\href {\doibase 10.1103/PhysRevX.9.031010}
  {\bibfield  {journal} {\bibinfo  {journal} {Phys. Rev. X}\ }\textbf {\bibinfo
  {volume} {9}},\ \bibinfo {pages} {031010} (\bibinfo {year}
  {2019})}\BibitemShut {NoStop}%
\bibitem [{\citenamefont {Chen}\ and\ \citenamefont {Xie}(2019)}]{XieXC2019}%
  \BibitemOpen
  \bibfield  {author} {\bibinfo {author} {\bibfnamefont {H.}~\bibnamefont
  {Chen}}\ and\ \bibinfo {author} {\bibfnamefont {X.~C.}\ \bibnamefont {Xie}},\
  }\href {\doibase 10.1103/PhysRevA.100.013601} {\bibfield  {journal} {\bibinfo
   {journal} {Phys. Rev. A}\ }\textbf {\bibinfo {volume} {100}},\ \bibinfo
  {pages} {013601} (\bibinfo {year} {2019})}\BibitemShut {NoStop}%
\bibitem [{\citenamefont {Li}\ \emph {et~al.}(2018)\citenamefont {Li},
  \citenamefont {Hanke}, \citenamefont {Hankiewicz}, \citenamefont {Reis},
  \citenamefont {Sch\"afer}, \citenamefont {Claessen}, \citenamefont {Wu},\
  and\ \citenamefont {Thomale}}]{Ligang2018}%
  \BibitemOpen
  \bibfield  {author} {\bibinfo {author} {\bibfnamefont {G.}~\bibnamefont
  {Li}}, \bibinfo {author} {\bibfnamefont {W.}~\bibnamefont {Hanke}}, \bibinfo
  {author} {\bibfnamefont {E.~M.}\ \bibnamefont {Hankiewicz}}, \bibinfo
  {author} {\bibfnamefont {F.}~\bibnamefont {Reis}}, \bibinfo {author}
  {\bibfnamefont {J.}~\bibnamefont {Sch\"afer}}, \bibinfo {author}
  {\bibfnamefont {R.}~\bibnamefont {Claessen}}, \bibinfo {author}
  {\bibfnamefont {C.}~\bibnamefont {Wu}}, \ and\ \bibinfo {author}
  {\bibfnamefont {R.}~\bibnamefont {Thomale}},\ }\href {\doibase
  10.1103/PhysRevB.98.165146} {\bibfield  {journal} {\bibinfo  {journal} {Phys.
  Rev. B}\ }\textbf {\bibinfo {volume} {98}},\ \bibinfo {pages} {165146}
  (\bibinfo {year} {2018})}\BibitemShut {NoStop}%
\bibitem [{\citenamefont {Canonico}\ \emph {et~al.}(2019)\citenamefont
  {Canonico}, \citenamefont {Rappoport},\ and\ \citenamefont
  {Muniz}}]{Canonico2019}%
  \BibitemOpen
  \bibfield  {author} {\bibinfo {author} {\bibfnamefont {L.~M.}\ \bibnamefont
  {Canonico}}, \bibinfo {author} {\bibfnamefont {T.~G.}\ \bibnamefont
  {Rappoport}}, \ and\ \bibinfo {author} {\bibfnamefont {R.~B.}\ \bibnamefont
  {Muniz}},\ }\href {\doibase 10.1103/PhysRevLett.122.196601} {\bibfield
  {journal} {\bibinfo  {journal} {Phys. Rev. Lett.}\ }\textbf {\bibinfo
  {volume} {122}},\ \bibinfo {pages} {196601} (\bibinfo {year}
  {2019})}\BibitemShut {NoStop}%
\bibitem [{\citenamefont {Lee}\ \emph {et~al.}(2010)\citenamefont {Lee},
  \citenamefont {Wu},\ and\ \citenamefont {Das~Sarma}}]{PhysRevA.82.053611}%
  \BibitemOpen
  \bibfield  {author} {\bibinfo {author} {\bibfnamefont {W.-C.}\ \bibnamefont
  {Lee}}, \bibinfo {author} {\bibfnamefont {C.}~\bibnamefont {Wu}}, \ and\
  \bibinfo {author} {\bibfnamefont {S.}~\bibnamefont {Das~Sarma}},\ }\href
  {\doibase 10.1103/PhysRevA.82.053611} {\bibfield  {journal} {\bibinfo
  {journal} {Phys. Rev. A}\ }\textbf {\bibinfo {volume} {82}},\ \bibinfo
  {pages} {053611} (\bibinfo {year} {2010})}\BibitemShut {NoStop}%
\bibitem [{\citenamefont {Zhou}\ \emph {et~al.}(2016)\citenamefont {Zhou},
  \citenamefont {Zhang}, \citenamefont {Xue}, \citenamefont {Zhao},
  \citenamefont {Zhang}, \citenamefont {Jiang},\ and\ \citenamefont
  {Yang}}]{zhou2016}%
  \BibitemOpen
  \bibfield  {author} {\bibinfo {author} {\bibfnamefont {T.}~\bibnamefont
  {Zhou}}, \bibinfo {author} {\bibfnamefont {J.}~\bibnamefont {Zhang}},
  \bibinfo {author} {\bibfnamefont {Y.}~\bibnamefont {Xue}}, \bibinfo {author}
  {\bibfnamefont {B.}~\bibnamefont {Zhao}}, \bibinfo {author} {\bibfnamefont
  {H.}~\bibnamefont {Zhang}}, \bibinfo {author} {\bibfnamefont
  {H.}~\bibnamefont {Jiang}}, \ and\ \bibinfo {author} {\bibfnamefont
  {Z.}~\bibnamefont {Yang}},\ }\href {\doibase 10.1103/PhysRevB.94.235449}
  {\bibfield  {journal} {\bibinfo  {journal} {Phys. Rev. B}\ }\textbf {\bibinfo
  {volume} {94}},\ \bibinfo {pages} {235449} (\bibinfo {year}
  {2016})}\BibitemShut {NoStop}%
\bibitem [{\citenamefont {Barreteau}\ \emph {et~al.}(2017)\citenamefont
  {Barreteau}, \citenamefont {Ducastelle},\ and\ \citenamefont
  {Mallah}}]{barreteau2017}%
  \BibitemOpen
  \bibfield  {author} {\bibinfo {author} {\bibfnamefont {C.}~\bibnamefont
  {Barreteau}}, \bibinfo {author} {\bibfnamefont {F.}~\bibnamefont
  {Ducastelle}}, \ and\ \bibinfo {author} {\bibfnamefont {T.}~\bibnamefont
  {Mallah}},\ }\href {http://stacks.iop.org/0953-8984/29/i=46/a=465302}
  {\bibfield  {journal} {\bibinfo  {journal} {J. Phys.: Condens. Matter}\
  }\textbf {\bibinfo {volume} {29}},\ \bibinfo {pages} {465302} (\bibinfo
  {year} {2017})}\BibitemShut {NoStop}%
\bibitem [{\citenamefont {Song}\ \emph {et~al.}(2018)\citenamefont {Song},
  \citenamefont {Yang}, \citenamefont {Du}, \citenamefont {Gao},\ and\
  \citenamefont {Yakobson}}]{song2018}%
  \BibitemOpen
  \bibfield  {author} {\bibinfo {author} {\bibfnamefont {S.-R.}\ \bibnamefont
  {Song}}, \bibinfo {author} {\bibfnamefont {J.-H.}\ \bibnamefont {Yang}},
  \bibinfo {author} {\bibfnamefont {S.-X.}\ \bibnamefont {Du}}, \bibinfo
  {author} {\bibfnamefont {H.-J.}\ \bibnamefont {Gao}}, \ and\ \bibinfo
  {author} {\bibfnamefont {B.~I.}\ \bibnamefont {Yakobson}},\ }\href {\doibase
  10.1088/1674-1056/27/8/087101} {\bibfield  {journal} {\bibinfo  {journal}
  {Chinese Phys. B}\ }\textbf {\bibinfo {volume} {27}},\ \bibinfo {pages}
  {087101} (\bibinfo {year} {2018})}\BibitemShut {NoStop}%
\bibitem [{\citenamefont {Jiang}\ \emph {et~al.}(2019)\citenamefont {Jiang},
  \citenamefont {Liu}, \citenamefont {Mei}, \citenamefont {Cui},\ and\
  \citenamefont {Liu}}]{jiang2019}%
  \BibitemOpen
  \bibfield  {author} {\bibinfo {author} {\bibfnamefont {W.}~\bibnamefont
  {Jiang}}, \bibinfo {author} {\bibfnamefont {Z.}~\bibnamefont {Liu}}, \bibinfo
  {author} {\bibfnamefont {J.-W.}\ \bibnamefont {Mei}}, \bibinfo {author}
  {\bibfnamefont {B.}~\bibnamefont {Cui}}, \ and\ \bibinfo {author}
  {\bibfnamefont {F.}~\bibnamefont {Liu}},\ }\href {\doibase
  10.1039/C8NR08479C} {\bibfield  {journal} {\bibinfo  {journal} {Nanoscale}\
  }\textbf {\bibinfo {volume} {11}},\ \bibinfo {pages} {955} (\bibinfo {year}
  {2019})}\BibitemShut {NoStop}%
\bibitem [{\citenamefont {Zhou}\ \emph {et~al.}(2019)\citenamefont {Zhou},
  \citenamefont {Zhou}, \citenamefont {Cheng}, \citenamefont {Jiang},\ and\
  \citenamefont {Yang}}]{zhou2019}%
  \BibitemOpen
  \bibfield  {author} {\bibinfo {author} {\bibfnamefont {J.}~\bibnamefont
  {Zhou}}, \bibinfo {author} {\bibfnamefont {T.}~\bibnamefont {Zhou}}, \bibinfo
  {author} {\bibfnamefont {S.-g.}\ \bibnamefont {Cheng}}, \bibinfo {author}
  {\bibfnamefont {H.}~\bibnamefont {Jiang}}, \ and\ \bibinfo {author}
  {\bibfnamefont {Z.}~\bibnamefont {Yang}},\ }\href {\doibase
  10.1103/PhysRevB.99.195422} {\bibfield  {journal} {\bibinfo  {journal} {Phys.
  Rev. B}\ }\textbf {\bibinfo {volume} {99}},\ \bibinfo {pages} {195422}
  (\bibinfo {year} {2019})}\BibitemShut {NoStop}%
\bibitem [{\citenamefont {Hu}\ \emph {et~al.}(2020)\citenamefont {Hu},
  \citenamefont {Wang},\ and\ \citenamefont {Li}}]{HU2020}%
  \BibitemOpen
  \bibfield  {author} {\bibinfo {author} {\bibfnamefont {X.-K.}\ \bibnamefont
  {Hu}}, \bibinfo {author} {\bibfnamefont {Y.}~\bibnamefont {Wang}}, \ and\
  \bibinfo {author} {\bibfnamefont {P.}~\bibnamefont {Li}},\ }\href {\doibase
  https://doi.org/10.1016/j.cplett.2019.137064} {\bibfield  {journal} {\bibinfo
   {journal} {Chem. Phys. Lett.}\ }\textbf {\bibinfo {volume} {740}},\ \bibinfo
  {pages} {137064} (\bibinfo {year} {2020})}\BibitemShut {NoStop}%
\bibitem [{\citenamefont {Jacqmin}\ \emph {et~al.}(2014)\citenamefont
  {Jacqmin}, \citenamefont {Carusotto}, \citenamefont {Sagnes}, \citenamefont
  {Abbarchi}, \citenamefont {Solnyshkov}, \citenamefont {Malpuech},
  \citenamefont {Galopin}, \citenamefont {Lema\^{\i}tre}, \citenamefont
  {Bloch},\ and\ \citenamefont {Amo}}]{amo2014}%
  \BibitemOpen
  \bibfield  {author} {\bibinfo {author} {\bibfnamefont {T.}~\bibnamefont
  {Jacqmin}}, \bibinfo {author} {\bibfnamefont {I.}~\bibnamefont {Carusotto}},
  \bibinfo {author} {\bibfnamefont {I.}~\bibnamefont {Sagnes}}, \bibinfo
  {author} {\bibfnamefont {M.}~\bibnamefont {Abbarchi}}, \bibinfo {author}
  {\bibfnamefont {D.~D.}\ \bibnamefont {Solnyshkov}}, \bibinfo {author}
  {\bibfnamefont {G.}~\bibnamefont {Malpuech}}, \bibinfo {author}
  {\bibfnamefont {E.}~\bibnamefont {Galopin}}, \bibinfo {author} {\bibfnamefont
  {A.}~\bibnamefont {Lema\^{\i}tre}}, \bibinfo {author} {\bibfnamefont
  {J.}~\bibnamefont {Bloch}}, \ and\ \bibinfo {author} {\bibfnamefont
  {A.}~\bibnamefont {Amo}},\ }\href {\doibase 10.1103/PhysRevLett.112.116402}
  {\bibfield  {journal} {\bibinfo  {journal} {Phys. Rev. Lett.}\ }\textbf
  {\bibinfo {volume} {112}},\ \bibinfo {pages} {116402} (\bibinfo {year}
  {2014})}\BibitemShut {NoStop}%
\bibitem [{\citenamefont {Ma}\ \emph {et~al.}(2019)\citenamefont {Ma},
  \citenamefont {Qiu}, \citenamefont {L\"u},\ and\ \citenamefont
  {Gao}}]{ma2017orbital}%
  \BibitemOpen
  \bibfield  {author} {\bibinfo {author} {\bibfnamefont {L.}~\bibnamefont
  {Ma}}, \bibinfo {author} {\bibfnamefont {W.-X.}\ \bibnamefont {Qiu}},
  \bibinfo {author} {\bibfnamefont {J.-T.}\ \bibnamefont {L\"u}}, \ and\
  \bibinfo {author} {\bibfnamefont {J.-H.}\ \bibnamefont {Gao}},\ }\href
  {\doibase 10.1103/PhysRevB.99.205403} {\bibfield  {journal} {\bibinfo
  {journal} {Phys. Rev. B}\ }\textbf {\bibinfo {volume} {99}},\ \bibinfo
  {pages} {205403} (\bibinfo {year} {2019})}\BibitemShut {NoStop}%
\bibitem [{\citenamefont {Slot}\ \emph {et~al.}(2017)\citenamefont {Slot},
  \citenamefont {Gardenier}, \citenamefont {Jacobse}, \citenamefont {van
  Miert}, \citenamefont {Kempkes}, \citenamefont {Zevenhuizen}, \citenamefont
  {Smith}, \citenamefont {Vanmaekelbergh},\ and\ \citenamefont
  {Swart}}]{slot2017experimental}%
  \BibitemOpen
  \bibfield  {author} {\bibinfo {author} {\bibfnamefont {M.~R.}\ \bibnamefont
  {Slot}}, \bibinfo {author} {\bibfnamefont {T.~S.}\ \bibnamefont {Gardenier}},
  \bibinfo {author} {\bibfnamefont {P.~H.}\ \bibnamefont {Jacobse}}, \bibinfo
  {author} {\bibfnamefont {G.~C.}\ \bibnamefont {van Miert}}, \bibinfo {author}
  {\bibfnamefont {S.~N.}\ \bibnamefont {Kempkes}}, \bibinfo {author}
  {\bibfnamefont {S.~J.}\ \bibnamefont {Zevenhuizen}}, \bibinfo {author}
  {\bibfnamefont {C.~M.}\ \bibnamefont {Smith}}, \bibinfo {author}
  {\bibfnamefont {D.}~\bibnamefont {Vanmaekelbergh}}, \ and\ \bibinfo {author}
  {\bibfnamefont {I.}~\bibnamefont {Swart}},\ }\href
  {https://doi.org/10.1038/nphys4105} {\bibfield  {journal} {\bibinfo
  {journal} {Nat. Phys.}\ }\textbf {\bibinfo {volume} {13}},\ \bibinfo {pages}
  {672} (\bibinfo {year} {2017})}\BibitemShut {NoStop}%
\bibitem [{\citenamefont {Gomes}\ \emph {et~al.}(2012)\citenamefont {Gomes},
  \citenamefont {Mar}, \citenamefont {Ko}, \citenamefont {Guinea},\ and\
  \citenamefont {Manoharan}}]{gomes2012designer}%
  \BibitemOpen
  \bibfield  {author} {\bibinfo {author} {\bibfnamefont {K.~K.}\ \bibnamefont
  {Gomes}}, \bibinfo {author} {\bibfnamefont {W.}~\bibnamefont {Mar}}, \bibinfo
  {author} {\bibfnamefont {W.}~\bibnamefont {Ko}}, \bibinfo {author}
  {\bibfnamefont {F.}~\bibnamefont {Guinea}}, \ and\ \bibinfo {author}
  {\bibfnamefont {H.~C.}\ \bibnamefont {Manoharan}},\ }\href
  {https://doi.org/10.1038/nature10941} {\bibfield  {journal} {\bibinfo
  {journal} {Nature}\ }\textbf {\bibinfo {volume} {483}},\ \bibinfo {pages}
  {306} (\bibinfo {year} {2012})}\BibitemShut {NoStop}%
\bibitem [{\citenamefont {Wang}\ \emph {et~al.}(2014)\citenamefont {Wang},
  \citenamefont {Tan}, \citenamefont {Wang}, \citenamefont {Louie},\ and\
  \citenamefont {Lin}}]{wang2014manipulation}%
  \BibitemOpen
  \bibfield  {author} {\bibinfo {author} {\bibfnamefont {S.}~\bibnamefont
  {Wang}}, \bibinfo {author} {\bibfnamefont {L.~Z.}\ \bibnamefont {Tan}},
  \bibinfo {author} {\bibfnamefont {W.}~\bibnamefont {Wang}}, \bibinfo {author}
  {\bibfnamefont {S.~G.}\ \bibnamefont {Louie}}, \ and\ \bibinfo {author}
  {\bibfnamefont {N.}~\bibnamefont {Lin}},\ }\href {\doibase
  10.1103/PhysRevLett.113.196803} {\bibfield  {journal} {\bibinfo  {journal}
  {Phys. Rev. Lett.}\ }\textbf {\bibinfo {volume} {113}},\ \bibinfo {pages}
  {196803} (\bibinfo {year} {2014})}\BibitemShut {NoStop}%
\bibitem [{\citenamefont {Park}\ and\ \citenamefont
  {Louie}(2009)}]{park2009making}%
  \BibitemOpen
  \bibfield  {author} {\bibinfo {author} {\bibfnamefont {C.-H.}\ \bibnamefont
  {Park}}\ and\ \bibinfo {author} {\bibfnamefont {S.~G.}\ \bibnamefont
  {Louie}},\ }\href {\doibase 10.1021/nl803706c} {\bibfield  {journal}
  {\bibinfo  {journal} {Nano Lett.}\ }\textbf {\bibinfo {volume} {9}},\
  \bibinfo {pages} {1793} (\bibinfo {year} {2009})}\BibitemShut {NoStop}%
\bibitem [{\citenamefont {Ropo}\ \emph
  {et~al.}(2014{\natexlab{a}})\citenamefont {Ropo}, \citenamefont
  {Paavilainen}, \citenamefont {Akola},\ and\ \citenamefont
  {R\"as\"anen}}]{ropo2014density}%
  \BibitemOpen
  \bibfield  {author} {\bibinfo {author} {\bibfnamefont {M.}~\bibnamefont
  {Ropo}}, \bibinfo {author} {\bibfnamefont {S.}~\bibnamefont {Paavilainen}},
  \bibinfo {author} {\bibfnamefont {J.}~\bibnamefont {Akola}}, \ and\ \bibinfo
  {author} {\bibfnamefont {E.}~\bibnamefont {R\"as\"anen}},\ }\href {\doibase
  10.1103/PhysRevB.90.241401} {\bibfield  {journal} {\bibinfo  {journal} {Phys.
  Rev. B}\ }\textbf {\bibinfo {volume} {90}},\ \bibinfo {pages} {241401}
  (\bibinfo {year} {2014}{\natexlab{a}})}\BibitemShut {NoStop}%
\bibitem [{\citenamefont {Paavilainen}\ \emph {et~al.}(2016)\citenamefont
  {Paavilainen}, \citenamefont {Ropo}, \citenamefont {Nieminen}, \citenamefont
  {Akola},\ and\ \citenamefont {R\"as\"anen}}]{Coexist2016}%
  \BibitemOpen
  \bibfield  {author} {\bibinfo {author} {\bibfnamefont {S.}~\bibnamefont
  {Paavilainen}}, \bibinfo {author} {\bibfnamefont {M.}~\bibnamefont {Ropo}},
  \bibinfo {author} {\bibfnamefont {J.}~\bibnamefont {Nieminen}}, \bibinfo
  {author} {\bibfnamefont {J.}~\bibnamefont {Akola}}, \ and\ \bibinfo {author}
  {\bibfnamefont {E.}~\bibnamefont {R\"as\"anen}},\ }\href
  {https://doi.org/10.1021/acs.nanolett.6b00397} {\bibfield  {journal}
  {\bibinfo  {journal} {Nano Lett.}\ }\textbf {\bibinfo {volume} {16}},\
  \bibinfo {pages} {3519} (\bibinfo {year} {2016})}\BibitemShut {NoStop}%
\bibitem [{\citenamefont {Li}\ \emph {et~al.}(2016)\citenamefont {Li},
  \citenamefont {Qiu},\ and\ \citenamefont {Gao}}]{li2016designing}%
  \BibitemOpen
  \bibfield  {author} {\bibinfo {author} {\bibfnamefont {S.}~\bibnamefont
  {Li}}, \bibinfo {author} {\bibfnamefont {W.-X.}\ \bibnamefont {Qiu}}, \ and\
  \bibinfo {author} {\bibfnamefont {J.-H.}\ \bibnamefont {Gao}},\ }\href
  {\doibase 10.1039/C6NR03223K} {\bibfield  {journal} {\bibinfo  {journal}
  {Nanoscale}\ }\textbf {\bibinfo {volume} {8}},\ \bibinfo {pages} {12747}
  (\bibinfo {year} {2016})}\BibitemShut {NoStop}%
\bibitem [{\citenamefont {Qiu}\ \emph {et~al.}(2016)\citenamefont {Qiu},
  \citenamefont {Li}, \citenamefont {Gao}, \citenamefont {Zhou},\ and\
  \citenamefont {Zhang}}]{qiu2016designing}%
  \BibitemOpen
  \bibfield  {author} {\bibinfo {author} {\bibfnamefont {W.-X.}\ \bibnamefont
  {Qiu}}, \bibinfo {author} {\bibfnamefont {S.}~\bibnamefont {Li}}, \bibinfo
  {author} {\bibfnamefont {J.-H.}\ \bibnamefont {Gao}}, \bibinfo {author}
  {\bibfnamefont {Y.}~\bibnamefont {Zhou}}, \ and\ \bibinfo {author}
  {\bibfnamefont {F.-C.}\ \bibnamefont {Zhang}},\ }\href {\doibase
  10.1103/PhysRevB.94.241409} {\bibfield  {journal} {\bibinfo  {journal} {Phys.
  Rev. B}\ }\textbf {\bibinfo {volume} {94}},\ \bibinfo {pages} {241409}
  (\bibinfo {year} {2016})}\BibitemShut {NoStop}%
\bibitem [{\citenamefont {Slot}\ \emph {et~al.}(2019)\citenamefont {Slot},
  \citenamefont {Kempkes}, \citenamefont {Knol}, \citenamefont {van
  Weerdenburg}, \citenamefont {van~den Broeke}, \citenamefont {Wegner},
  \citenamefont {Vanmaekelbergh}, \citenamefont {Khajetoorians}, \citenamefont
  {Morais~Smith},\ and\ \citenamefont {Swart}}]{slot2018experimental}%
  \BibitemOpen
  \bibfield  {author} {\bibinfo {author} {\bibfnamefont {M.~R.}\ \bibnamefont
  {Slot}}, \bibinfo {author} {\bibfnamefont {S.~N.}\ \bibnamefont {Kempkes}},
  \bibinfo {author} {\bibfnamefont {E.~J.}\ \bibnamefont {Knol}}, \bibinfo
  {author} {\bibfnamefont {W.~M.~J.}\ \bibnamefont {van Weerdenburg}}, \bibinfo
  {author} {\bibfnamefont {J.~J.}\ \bibnamefont {van~den Broeke}}, \bibinfo
  {author} {\bibfnamefont {D.}~\bibnamefont {Wegner}}, \bibinfo {author}
  {\bibfnamefont {D.}~\bibnamefont {Vanmaekelbergh}}, \bibinfo {author}
  {\bibfnamefont {A.~A.}\ \bibnamefont {Khajetoorians}}, \bibinfo {author}
  {\bibfnamefont {C.}~\bibnamefont {Morais~Smith}}, \ and\ \bibinfo {author}
  {\bibfnamefont {I.}~\bibnamefont {Swart}},\ }\href {\doibase
  10.1103/PhysRevX.9.011009} {\bibfield  {journal} {\bibinfo  {journal} {Phys.
  Rev. X}\ }\textbf {\bibinfo {volume} {9}},\ \bibinfo {pages} {011009}
  (\bibinfo {year} {2019})}\BibitemShut {NoStop}%
\bibitem [{\citenamefont {Slater}\ and\ \citenamefont
  {Koster}(1954)}]{Slater-Koster}%
  \BibitemOpen
  \bibfield  {author} {\bibinfo {author} {\bibfnamefont {J.~C.}\ \bibnamefont
  {Slater}}\ and\ \bibinfo {author} {\bibfnamefont {G.~F.}\ \bibnamefont
  {Koster}},\ }\href {\doibase 10.1103/PhysRev.94.1498} {\bibfield  {journal}
  {\bibinfo  {journal} {Phys. Rev.}\ }\textbf {\bibinfo {volume} {94}},\
  \bibinfo {pages} {1498} (\bibinfo {year} {1954})}\BibitemShut {NoStop}%
\bibitem [{\citenamefont {Ropo}\ \emph
  {et~al.}(2014{\natexlab{b}})\citenamefont {Ropo}, \citenamefont
  {Paavilainen}, \citenamefont {Akola},\ and\ \citenamefont
  {R\"as\"anen}}]{dft2014}%
  \BibitemOpen
  \bibfield  {author} {\bibinfo {author} {\bibfnamefont {M.}~\bibnamefont
  {Ropo}}, \bibinfo {author} {\bibfnamefont {S.}~\bibnamefont {Paavilainen}},
  \bibinfo {author} {\bibfnamefont {J.}~\bibnamefont {Akola}}, \ and\ \bibinfo
  {author} {\bibfnamefont {E.}~\bibnamefont {R\"as\"anen}},\ }\href {\doibase
  10.1103/PhysRevB.90.241401} {\bibfield  {journal} {\bibinfo  {journal} {Phys.
  Rev. B}\ }\textbf {\bibinfo {volume} {90}},\ \bibinfo {pages} {241401}
  (\bibinfo {year} {2014}{\natexlab{b}})}\BibitemShut {NoStop}%
\bibitem [{\citenamefont {Kresse}\ and\ \citenamefont
  {Furthm\"uller}(1996)}]{vasp}%
  \BibitemOpen
  \bibfield  {author} {\bibinfo {author} {\bibfnamefont {G.}~\bibnamefont
  {Kresse}}\ and\ \bibinfo {author} {\bibfnamefont {J.}~\bibnamefont
  {Furthm\"uller}},\ }\href {\doibase 10.1103/PhysRevB.54.11169} {\bibfield
  {journal} {\bibinfo  {journal} {Phys. Rev. B}\ }\textbf {\bibinfo {volume}
  {54}},\ \bibinfo {pages} {11169} (\bibinfo {year} {1996})}\BibitemShut
  {NoStop}%
\bibitem [{\citenamefont {Perdew}\ \emph {et~al.}(1996)\citenamefont {Perdew},
  \citenamefont {Burke},\ and\ \citenamefont {Ernzerhof}}]{PBE}%
  \BibitemOpen
  \bibfield  {author} {\bibinfo {author} {\bibfnamefont {J.~P.}\ \bibnamefont
  {Perdew}}, \bibinfo {author} {\bibfnamefont {K.}~\bibnamefont {Burke}}, \
  and\ \bibinfo {author} {\bibfnamefont {M.}~\bibnamefont {Ernzerhof}},\ }\href
  {\doibase 10.1103/PhysRevLett.77.3865} {\bibfield  {journal} {\bibinfo
  {journal} {Phys. Rev. Lett.}\ }\textbf {\bibinfo {volume} {77}},\ \bibinfo
  {pages} {3865} (\bibinfo {year} {1996})}\BibitemShut {NoStop}%
\bibitem [{\citenamefont {Tersoff}\ and\ \citenamefont
  {Hamann}(1983)}]{tersoff1983}%
  \BibitemOpen
  \bibfield  {author} {\bibinfo {author} {\bibfnamefont {J.}~\bibnamefont
  {Tersoff}}\ and\ \bibinfo {author} {\bibfnamefont {D.~R.}\ \bibnamefont
  {Hamann}},\ }\href {\doibase 10.1103/PhysRevLett.50.1998} {\bibfield
  {journal} {\bibinfo  {journal} {Phys. Rev. Lett.}\ }\textbf {\bibinfo
  {volume} {50}},\ \bibinfo {pages} {1998} (\bibinfo {year}
  {1983})}\BibitemShut {NoStop}%
\bibitem [{\citenamefont {{Wang}}\ \emph {et~al.}(2019)\citenamefont {{Wang}},
  \citenamefont {{Xu}}, \citenamefont {{Liu}}, \citenamefont {{Tang}},\ and\
  \citenamefont {{Geng}}}]{vei2019}%
  \BibitemOpen
  \bibfield  {author} {\bibinfo {author} {\bibfnamefont {V.}~\bibnamefont
  {{Wang}}}, \bibinfo {author} {\bibfnamefont {N.}~\bibnamefont {{Xu}}},
  \bibinfo {author} {\bibfnamefont {J.~C.}\ \bibnamefont {{Liu}}}, \bibinfo
  {author} {\bibfnamefont {G.}~\bibnamefont {{Tang}}}, \ and\ \bibinfo {author}
  {\bibfnamefont {W.-T.}\ \bibnamefont {{Geng}}},\ }\href@noop {} {\bibfield
  {journal} {\bibinfo  {journal} {arXiv e-prints}\ ,\ \bibinfo {eid}
  {arXiv:1908.08269}} (\bibinfo {year} {2019})},\ \Eprint
  {http://arxiv.org/abs/1908.08269} {arXiv:1908.08269 [cond-mat.mtrl-sci]}
  \BibitemShut {NoStop}%
\bibitem [{\citenamefont {Butti}\ \emph {et~al.}(2005)\citenamefont {Butti},
  \citenamefont {Caravati}, \citenamefont {Brivio}, \citenamefont {Trioni},\
  and\ \citenamefont {Ishida}}]{butti2005image}%
  \BibitemOpen
  \bibfield  {author} {\bibinfo {author} {\bibfnamefont {G.}~\bibnamefont
  {Butti}}, \bibinfo {author} {\bibfnamefont {S.}~\bibnamefont {Caravati}},
  \bibinfo {author} {\bibfnamefont {G.~P.}\ \bibnamefont {Brivio}}, \bibinfo
  {author} {\bibfnamefont {M.~I.}\ \bibnamefont {Trioni}}, \ and\ \bibinfo
  {author} {\bibfnamefont {H.}~\bibnamefont {Ishida}},\ }\href {\doibase
  10.1103/PhysRevB.72.125402} {\bibfield  {journal} {\bibinfo  {journal} {Phys.
  Rev. B}\ }\textbf {\bibinfo {volume} {72}},\ \bibinfo {pages} {125402}
  (\bibinfo {year} {2005})}\BibitemShut {NoStop}%
\bibitem [{\citenamefont {Berland}\ \emph {et~al.}(2012)\citenamefont
  {Berland}, \citenamefont {Einstein},\ and\ \citenamefont
  {Hyldgaard}}]{berland2012response}%
  \BibitemOpen
  \bibfield  {author} {\bibinfo {author} {\bibfnamefont {K.}~\bibnamefont
  {Berland}}, \bibinfo {author} {\bibfnamefont {T.~L.}\ \bibnamefont
  {Einstein}}, \ and\ \bibinfo {author} {\bibfnamefont {P.}~\bibnamefont
  {Hyldgaard}},\ }\href {\doibase 10.1103/PhysRevB.85.035427} {\bibfield
  {journal} {\bibinfo  {journal} {Phys. Rev. B}\ }\textbf {\bibinfo {volume}
  {85}},\ \bibinfo {pages} {035427} (\bibinfo {year} {2012})}\BibitemShut
  {NoStop}%
\bibitem [{\citenamefont {Courths}\ \emph {et~al.}(2001)\citenamefont
  {Courths}, \citenamefont {Lau}, \citenamefont {Scheunemann}, \citenamefont
  {Gollisch},\ and\ \citenamefont {Feder}}]{courths2001from}%
  \BibitemOpen
  \bibfield  {author} {\bibinfo {author} {\bibfnamefont {R.}~\bibnamefont
  {Courths}}, \bibinfo {author} {\bibfnamefont {M.}~\bibnamefont {Lau}},
  \bibinfo {author} {\bibfnamefont {T.}~\bibnamefont {Scheunemann}}, \bibinfo
  {author} {\bibfnamefont {H.}~\bibnamefont {Gollisch}}, \ and\ \bibinfo
  {author} {\bibfnamefont {R.}~\bibnamefont {Feder}},\ }\href {\doibase
  10.1103/PhysRevB.63.195110} {\bibfield  {journal} {\bibinfo  {journal} {Phys.
  Rev. B}\ }\textbf {\bibinfo {volume} {63}},\ \bibinfo {pages} {195110}
  (\bibinfo {year} {2001})}\BibitemShut {NoStop}%
\bibitem [{\citenamefont {Kevan}(1983)}]{kevan1983evidence}%
  \BibitemOpen
  \bibfield  {author} {\bibinfo {author} {\bibfnamefont {S.~D.}\ \bibnamefont
  {Kevan}},\ }\href {\doibase 10.1103/PhysRevLett.50.526} {\bibfield  {journal}
  {\bibinfo  {journal} {Phys. Rev. Lett.}\ }\textbf {\bibinfo {volume} {50}},\
  \bibinfo {pages} {526} (\bibinfo {year} {1983})}\BibitemShut {NoStop}%
\bibitem [{\citenamefont {Jeandupeux}\ \emph {et~al.}(1999)\citenamefont
  {Jeandupeux}, \citenamefont {B\"urgi}, \citenamefont {Hirstein},
  \citenamefont {Brune},\ and\ \citenamefont {Kern}}]{jeandupeux1999thermal}%
  \BibitemOpen
  \bibfield  {author} {\bibinfo {author} {\bibfnamefont {O.}~\bibnamefont
  {Jeandupeux}}, \bibinfo {author} {\bibfnamefont {L.}~\bibnamefont {B\"urgi}},
  \bibinfo {author} {\bibfnamefont {A.}~\bibnamefont {Hirstein}}, \bibinfo
  {author} {\bibfnamefont {H.}~\bibnamefont {Brune}}, \ and\ \bibinfo {author}
  {\bibfnamefont {K.}~\bibnamefont {Kern}},\ }\href {\doibase
  10.1103/PhysRevB.59.15926} {\bibfield  {journal} {\bibinfo  {journal} {Phys.
  Rev. B}\ }\textbf {\bibinfo {volume} {59}},\ \bibinfo {pages} {15926}
  (\bibinfo {year} {1999})}\BibitemShut {NoStop}%
\bibitem [{\citenamefont {Tamai}\ \emph {et~al.}(2013)\citenamefont {Tamai},
  \citenamefont {Meevasana}, \citenamefont {King}, \citenamefont {Nicholson},
  \citenamefont {de~la Torre}, \citenamefont {Rozbicki},\ and\ \citenamefont
  {Baumberger}}]{tamai2013spin}%
  \BibitemOpen
  \bibfield  {author} {\bibinfo {author} {\bibfnamefont {A.}~\bibnamefont
  {Tamai}}, \bibinfo {author} {\bibfnamefont {W.}~\bibnamefont {Meevasana}},
  \bibinfo {author} {\bibfnamefont {P.~D.~C.}\ \bibnamefont {King}}, \bibinfo
  {author} {\bibfnamefont {C.~W.}\ \bibnamefont {Nicholson}}, \bibinfo {author}
  {\bibfnamefont {A.}~\bibnamefont {de~la Torre}}, \bibinfo {author}
  {\bibfnamefont {E.}~\bibnamefont {Rozbicki}}, \ and\ \bibinfo {author}
  {\bibfnamefont {F.}~\bibnamefont {Baumberger}},\ }\href {\doibase
  10.1103/PhysRevB.87.075113} {\bibfield  {journal} {\bibinfo  {journal} {Phys.
  Rev. B}\ }\textbf {\bibinfo {volume} {87}},\ \bibinfo {pages} {075113}
  (\bibinfo {year} {2013})}\BibitemShut {NoStop}%
\bibitem [{\citenamefont {N{\'{a}}dvorn{\'{\i}}k}\ \emph
  {et~al.}(2012)\citenamefont {N{\'{a}}dvorn{\'{\i}}k}, \citenamefont {Orlita},
  \citenamefont {Goncharuk}, \citenamefont {Smr{\v{c}}ka}, \citenamefont
  {Nov{\'{a}}k}, \citenamefont {Jurka}, \citenamefont {Hru{\v{s}}ka},
  \citenamefont {V{\'{y}}born{\'{y}}}, \citenamefont {Wasilewski},
  \citenamefont {Potemski},\ and\ \citenamefont
  {V{\'{y}}born{\'{y}}}}]{dvorn2012}%
  \BibitemOpen
  \bibfield  {author} {\bibinfo {author} {\bibfnamefont {L.}~\bibnamefont
  {N{\'{a}}dvorn{\'{\i}}k}}, \bibinfo {author} {\bibfnamefont {M.}~\bibnamefont
  {Orlita}}, \bibinfo {author} {\bibfnamefont {N.~A.}\ \bibnamefont
  {Goncharuk}}, \bibinfo {author} {\bibfnamefont {L.}~\bibnamefont
  {Smr{\v{c}}ka}}, \bibinfo {author} {\bibfnamefont {V.}~\bibnamefont
  {Nov{\'{a}}k}}, \bibinfo {author} {\bibfnamefont {V.}~\bibnamefont {Jurka}},
  \bibinfo {author} {\bibfnamefont {K.}~\bibnamefont {Hru{\v{s}}ka}}, \bibinfo
  {author} {\bibfnamefont {Z.}~\bibnamefont {V{\'{y}}born{\'{y}}}}, \bibinfo
  {author} {\bibfnamefont {Z.~R.}\ \bibnamefont {Wasilewski}}, \bibinfo
  {author} {\bibfnamefont {M.}~\bibnamefont {Potemski}}, \ and\ \bibinfo
  {author} {\bibfnamefont {K.}~\bibnamefont {V{\'{y}}born{\'{y}}}},\ }\href
  {\doibase 10.1088/1367-2630/14/5/053002} {\bibfield  {journal} {\bibinfo
  {journal} {New Journal of Physics}\ }\textbf {\bibinfo {volume} {14}},\
  \bibinfo {pages} {053002} (\bibinfo {year} {2012})}\BibitemShut {NoStop}%
\bibitem [{\citenamefont {Polini}\ \emph {et~al.}(2013)\citenamefont {Polini},
  \citenamefont {Guinea}, \citenamefont {Lewenstein}, \citenamefont
  {Manoharan},\ and\ \citenamefont {Pellegrini}}]{polini2013artificial}%
  \BibitemOpen
  \bibfield  {author} {\bibinfo {author} {\bibfnamefont {M.}~\bibnamefont
  {Polini}}, \bibinfo {author} {\bibfnamefont {F.}~\bibnamefont {Guinea}},
  \bibinfo {author} {\bibfnamefont {M.}~\bibnamefont {Lewenstein}}, \bibinfo
  {author} {\bibfnamefont {H.~C.}\ \bibnamefont {Manoharan}}, \ and\ \bibinfo
  {author} {\bibfnamefont {V.}~\bibnamefont {Pellegrini}},\ }\href
  {https://doi.org/10.1038/nnano.2013.161} {\bibfield  {journal} {\bibinfo
  {journal} {Nature nanotechnology}\ }\textbf {\bibinfo {volume} {8}},\
  \bibinfo {pages} {625} (\bibinfo {year} {2013})}\BibitemShut {NoStop}%
\bibitem [{\citenamefont {Klembt}\ \emph {et~al.}(2018)\citenamefont {Klembt},
  \citenamefont {Harder}, \citenamefont {Egorov}, \citenamefont {Winkler},
  \citenamefont {Ge}, \citenamefont {Bandres}, \citenamefont {Emmerling},
  \citenamefont {Worschech}, \citenamefont {Liew}, \citenamefont {Segev},
  \citenamefont {Schneider},\ and\ \citenamefont
  {H$\ddot{o}$fling}}]{klembt2018}%
  \BibitemOpen
  \bibfield  {author} {\bibinfo {author} {\bibfnamefont {S.}~\bibnamefont
  {Klembt}}, \bibinfo {author} {\bibfnamefont {T.~H.}\ \bibnamefont {Harder}},
  \bibinfo {author} {\bibfnamefont {O.~A.}\ \bibnamefont {Egorov}}, \bibinfo
  {author} {\bibfnamefont {K.}~\bibnamefont {Winkler}}, \bibinfo {author}
  {\bibfnamefont {R.}~\bibnamefont {Ge}}, \bibinfo {author} {\bibfnamefont
  {M.~A.}\ \bibnamefont {Bandres}}, \bibinfo {author} {\bibfnamefont
  {M.}~\bibnamefont {Emmerling}}, \bibinfo {author} {\bibfnamefont
  {L.}~\bibnamefont {Worschech}}, \bibinfo {author} {\bibfnamefont {T.~C.~H.}\
  \bibnamefont {Liew}}, \bibinfo {author} {\bibfnamefont {M.}~\bibnamefont
  {Segev}}, \bibinfo {author} {\bibfnamefont {C.}~\bibnamefont {Schneider}}, \
  and\ \bibinfo {author} {\bibfnamefont {S.}~\bibnamefont {H$\ddot{o}$fling}},\
  }\href {\doibase 10.1038/s41586-018-0601-5} {\bibfield  {journal} {\bibinfo
  {journal} {Nature}\ }\textbf {\bibinfo {volume} {562}},\ \bibinfo {pages}
  {552} (\bibinfo {year} {2018})}\BibitemShut {NoStop}%
\bibitem [{\citenamefont {{Qiu}}\ \emph {et~al.}(2019)\citenamefont {{Qiu}},
  \citenamefont {{Ma}}, \citenamefont {{L{\"u}}},\ and\ \citenamefont
  {{Gao}}}]{qiu2019making}%
  \BibitemOpen
  \bibfield  {author} {\bibinfo {author} {\bibfnamefont {W.-X.}\ \bibnamefont
  {{Qiu}}}, \bibinfo {author} {\bibfnamefont {L.}~\bibnamefont {{Ma}}},
  \bibinfo {author} {\bibfnamefont {J.-T.}\ \bibnamefont {{L{\"u}}}}, \ and\
  \bibinfo {author} {\bibfnamefont {J.-H.}\ \bibnamefont {{Gao}}},\ }\href@noop
  {} {\bibfield  {journal} {\bibinfo  {journal} {arXiv e-prints}\ ,\ \bibinfo
  {eid} {arXiv:1901.01008}} (\bibinfo {year} {2019})},\ \Eprint
  {http://arxiv.org/abs/1901.01008} {arXiv:1901.01008 [cond-mat.mes-hall]}
  \BibitemShut {NoStop}%
\bibitem [{\citenamefont {Gardenier}\ \emph {et~al.}(2020)\citenamefont
  {Gardenier}, \citenamefont {van~den Broeke}, \citenamefont {Moes},
  \citenamefont {Swart}, \citenamefont {Delerue}, \citenamefont {Slot},
  \citenamefont {Smith},\ and\ \citenamefont {Vanmaekelbergh}}]{gardenier2020}%
  \BibitemOpen
  \bibfield  {author} {\bibinfo {author} {\bibfnamefont {T.~S.}\ \bibnamefont
  {Gardenier}}, \bibinfo {author} {\bibfnamefont {J.~J.}\ \bibnamefont {van~den
  Broeke}}, \bibinfo {author} {\bibfnamefont {J.~R.}\ \bibnamefont {Moes}},
  \bibinfo {author} {\bibfnamefont {I.}~\bibnamefont {Swart}}, \bibinfo
  {author} {\bibfnamefont {C.}~\bibnamefont {Delerue}}, \bibinfo {author}
  {\bibfnamefont {M.~R.}\ \bibnamefont {Slot}}, \bibinfo {author}
  {\bibfnamefont {C.~M.}\ \bibnamefont {Smith}}, \ and\ \bibinfo {author}
  {\bibfnamefont {D.}~\bibnamefont {Vanmaekelbergh}},\ }\href {\doibase
  10.1021/acsnano.0c05747} {\bibfield  {journal} {\bibinfo  {journal} {ACS
  Nano}\ }\textbf {\bibinfo {volume} {14}},\ \bibinfo {pages} {13638} (\bibinfo
  {year} {2020})},\ \bibinfo {note} {pMID: 32991147}\BibitemShut {NoStop}%
\end{thebibliography}
\end{document}